\documentclass[12pt,a4paper]{article}
\usepackage{graphicx,epsf, amsfonts}
\title{Polarized Neutron Scattering}
\author{B. Roessli$^1$, P. B\"oni$^2$ \\\small{$^1$Laboratory for Neutron
Scattering, ETH Z\"urich \& Paul Scherrer Institute, CH-5232
Villigen PSI}\\\small{$^2$ Physik-Department E21, Technische
Universit\"at M\"unchen, D-85748 Garching }}

\date{\today}

\begin{document}

\maketitle

%\begin{abstract}

%\indent

%\end{abstract}


\tableofcontents

\newpage

\section{Introduction}

In the previous chapters of this book, experimental results and
the theory of elastic and inelastic neutron scattering have been
presented under the assumption that the magnetic moments of the
neutrons are randomly oriented. It has been shown that details
about the physical properties of a system are extracted by
analysing the momentum and the energy of the scattered neutrons.
It is intuitively imaginable, however, that measuring the spin
state of the neutron after scattering relative to its state before
the scattering process should provide us with additional
information. To that end the cross-sections for neutron scattering
must now also take into account the relationship between the spin
of the neutron  with the physical properties of the target.

The polarization of a neutron is defined as
\begin{equation}
\mathbb{P}=2\langle \mathbf{\hat s} \rangle= \langle \mbox{\boldmath$\hat \sigma$}\rangle,
\end{equation}
where $ \mbox{\boldmath$\sigma$}$ are the Pauli matrices. Clearly,
$|\mathbb{P}|$ is equal to $0$ for a completely unpolarised beam
and $|\mathbb{P}|=1$ if the beam is totally polarised. For
intermediate values, the neutron beam is not in a well defined
state and the spin part of the neutron wave-function must be
described by a more general form $\chi=u\chi_\uparrow
+v\chi_\downarrow$ with $|u|^2+|v|^2=1$. That is $|u|^2$ and
$|v|^2$ are the probabilities that the neutron spin will be
\textit{up} or \textit{down}, respectively. If a matrix operator
$\rm \hat \rho$ is defined like
\begin{equation}
\hat \rho= \chi\chi^\dagger = \left(
\begin{array}{cc}
|u|^2 & uv^\dagger \\
vu^\dagger & |v|^2
\end{array}
\right)={1\over 2}\left(\mathbb{I}+\mathbf{P}\cdot
\mbox{$\mathbf{\hat \sigma}$}\right) ,
\end{equation}
then the polarization of the neutron is described by a
three-dimensional \textit{vector} wit components $\mathbf
P=(2\Re(u^\dagger v), 2\Im(u^\dagger v), |u|^2-|v|^2)$
\cite{lovesey}. The polarization of a neutron beam is accordingly
defined as $\mathbf{P}={1\over N}\sum \mathbf{P}_j$ where $N$ is
the total number of neutrons and the sum runs over the
polarization vector of the individual neutrons $j$.

The cross-section $\sigma$ and the polarization of the scattered
beam $\mathbf{P}_f$ can be expressed as a function of the density
matrix $\hat \rho$, the polarization vector of the incident
neutrons $\mathbf{P}_i$ and the interaction potential $\hat v$
between the target and the neutron. In its most general form,
$\mathbf{P}_f=Tr\hat\rho\hat v^\dagger \mbox{\boldmath$\hat
\sigma$}\hat v/Tr\hat\rho\hat v^\dagger\hat v$.

Neutron scattering with polarized neutrons has been used in
fundamental and condensed matter physics since many years despite
the low flux of polarized beams. One of the first applications was
the study of spin density distributions in ferromagnets, following
the pioneering work of Shull and Nathans \cite{nath}. In those
days the polarization of the scattered neutrons was established by
measuring their transmission through a magnetized block of iron.
As predicted by Halpern and Johnson \cite{halp} the polarization
of the scattered neutrons, $\mathbf P_f$, depends on the
orientation of the scattering vector $\mathbf Q$ with respect to
the polarization of the incident neutrons, $\mathbf P_i$, like
\begin{equation}
\mathbf P_f = - \hat Q (\hat Q \cdot \mathbf P_i) \label{pola}
\end{equation}
where $\hat Q = \mathbf Q/|\mathbf Q|$.  In other words for
$\mathbf P_i
\parallel \mathbf Q$ all magnetic scattering is spin flip. Therefore
polarization analysis in neutron scattering provides an excellent
method to distinguish between nuclear and magnetic scattering. In
1969 the classic paper by Moon et al. \cite{moon1} appeared that
explains in simple terms
one-dimensional polarization analysis (nowadays called longitudinal
polarimetry) of
neutrons for elastic as well as inelastic neutron scattering.
They demonstrated the polarization dependence of the nuclear and
magnetic scattering.

Another, rather different application of polarized neutrons is
their use for attaining extremely high energy resolution by
measuring changes in the neutron beam polarization caused by
in-elastic scattering.  In 1972, Mezei \cite{meze} suggested to
use the precession of the magnetic moment of the neutrons in a
magnetic field as an internal clock. By means of the so-called
\textit{neutron spin-echo} technique energy resolutions of the
order of nano-electron volts can be achieved enabling the
investigation of slow dynamics, for example in the critical
region of magnetic systems or in polymers and glasses.

Nowadays, polarized neutron scattering is a fast developing
experimental method that finds applications in various fields of
condensed-matter research. Examples are
\begin{itemize}
\item{ determination of magnetic structures and spin densities,}
\item{ identification of magnetic fluctuations and their different
modes,}
%%%%%%%%%%%%%%%%%%%%%%%%%%%%%%\item{ comprehensive investigation of spin-phonon coupling}
\item{ separation of coherent from incoherent processes.}
\end{itemize}

In the following, necessarily incomplete sections, we shall
provide a presentation of the polarization dependence of neutron
cross-sections and show how the different scattering processes can
influence the polarization of the neutron beams. We will then
explain how polarized neutron beams can be produced and the
polarization determined after scattering. Finally we shall give
examples, where the technique of polarized neutron scattering can
provide new insight into physical processes in condensed matter
research. For a more detailed introduction into the field of
polarized neutron scattering with refer the interested reader to
the book of Williams \cite{williams}.

\section{Elastic neutron scattering cross-section for polarized neutrons}

The theory of elastic neutron scattering taking into account
polarization effects has been derived by Blume in 1963
\cite{blume}. The complete description of the scattering process
involving both nuclear and magnetic interactions can be given by
means of two master equations. The first one gives the total
neutron cross-section which depends on the polarization
$\mathbf P_i$ of the incident neutron beam as follows:
 \begin{equation}
 \sigma=NN^*+\mathbf D_\perp \cdot \mathbf D_\perp^* +
 \mathbf P_i (\mathbf D_\perp N^* + \mathbf D_\perp^*N) +i\mathbf P_i (\mathbf D_\perp^*
\times \mathbf D_\perp),
 \label{cross-section}
 \end{equation}
where $\sigma$ is the total cross section expressed in $\rm
[cm^2]$. For simplicity, the contribution of the nuclear spins is
neglected. $N=N(\mathbf Q)=\sum_i {b_i\exp{(i\mathbf Q \cdot \mathbf
r_i)}}$ is the structure factor of the atomic structure that
depends on the scattering vector $\mathbf Q$ and the scattering
lengths of the individual nuclei $b_i$; $\mathbf D_\perp$ is the
magnetic interaction vector with $\mathbf D_\perp = \mathbf
D_\perp (\mathbf Q )= \hat Q \times (\mathbf \rho (\mathbf Q)
\times \hat Q)$. $\mathbf \rho (\mathbf Q)$ is the Fourier
transform of the magnetic moment distribution and $\hat Q={{\mathbf Q} /
{|\mathbf Q|}}$. Therefore, only magnetic components perpendicular
to the scattering vector participate in the scattering process.
The scalar of the polarization vector $\mathbf P_i$ reflects the
degree of polarization of the neutrons, being equal to $\pm 1$ for
a fully polarized beam.

Eq. \ref{cross-section} shows that the neutron cross-section
depends only on the square of the chemical and magnetic structure
factor if a non-polarized neutron beam ($\mathbf P_i=0$) is used.
For a fully polarized beam ($|\mathbf P_i|=1$), two additional
terms contribute to the scattering, namely the
\textit{magnetic-nuclear interference} term and the
\textit{chiral} term, respectively. The magnetic-nuclear
interference term being proportional to $\mathbf D_\perp N^* +
\mathbf D_\perp^* N$ yields only a non-vanishing contribution to
the neutron cross-section if a Bragg reflection is due to nuclear
and magnetic scattering, like in ferromagnets and in
non-centrosymmetric antiferromagnets with propagation vector
$\mathbf Q_0 = 0$. The chiral term $\mathbf D_\perp^* \times \mathbf
D_\perp$ is non-zero whenever $\mathbf D_\perp$ is not parallel to
$\mathbf D_\perp^*$, as it is the case e.g. for a helicoidal
magnetic structure.

The second master equation provides the polarization of the
neutron beam \textit{after} the scattering process relative to the
polarization of the incident neutron beam:
\begin{eqnarray}
\mathbf P_f \sigma & = &\mathbf P_i NN^* \nonumber \\
    & + &(-1)\mathbf P_i(\mathbf D_\perp\cdot\mathbf D_\perp^*)+\mathbf D_\perp (\mathbf P_i\cdot\mathbf D_\perp^*)+\mathbf D_\perp^* (\mathbf P_i\cdot\mathbf D_\perp) \nonumber \\
                  & + &\mathbf D_\perp N^* + \mathbf D_\perp^* N +i(\mathbf D_\perp N^* - \mathbf
D_\perp^* N)\times \mathbf P_i \nonumber \\
                  & + &i\mathbf D_\perp\times\mathbf D_\perp^* ,
\label{polar}
\end{eqnarray}
where $\mathbf P_f$ is the polarization vector of the scattered
neutrons. On the one hand, Eq. \ref{polar} shows that pure nuclear
scattering ($\mathbf D_\perp=0$) leaves the polarization of the
neutron beam unchanged. On the other hand, polarisation of the scattered  beam
is obtained either, as we will see in the next chapter, by scattering neutrons
on mixed nuclear-magnetic Bragg reflections or from a helicoidal magnetic structure.
In the latter case, with $\rm \mathbf D_\perp N^* + \mathbf D_\perp^* N=0$,
a polarized beam with a polarization given by
\begin{equation}
\mathbf P_f={{i\mathbf D_\perp\times\mathbf
D_\perp^*}\over{\sigma}}= {{i\mathbf D_\perp\times\mathbf
D_\perp^*}\over{\mathbf D_\perp \cdot \mathbf D_\perp^* }}
\end{equation}
is created.
We point out that a measurement of the chiral term provides the
helicity of a helicoidal magnetic structure as has been shown by
Shirane et al. \cite{shirane}. In the general case, the
polarization vector of the neutron beam after scattering is
rotated with respect to $\mathbf P_i$ and its length is not
necessarily equal to $|\mathbf P_i|$. The term
\textit{``polarization analysis''} therefore refers to the
determination of the direction and length of $\mathbf P_f$.

\section{Production of polarized neutrons}

For a measurement of the polarization dependence of cross sections
various techniques to produce and analyze polarized neutron beams
have been developed. Depending on the required phase space
properties of the beams, i.e. continuous vs. pulsed, energy,
divergence, type of detector, etc., different methods for the spin
analysis are used. The most common methods are diffraction from
single-crystal polarizers (mostly Heusler), reflection from
magnetized thin film multi-layers or supermirrors, and absorption
of the non-wanted spin state by means of polarized $^3$He. A
recent review of these techniques can be found in Ref.
\cite{anderson}.

\subsection{Single-crystal polarizers}

This method produces a polarised neutron beam by taking advantage of the magnetic-nuclear
interference term in Eq. \ref{cross-section} and Eq. \ref{polar}.
If a magnetic field is applied to a centro-symmetric crystal so
that all the magnetic moments are saturated and aligned
perpendicular to the scattering vector $\mathbf Q$, the neutron
scattering cross-section for Bragg scattering is given by (set
$\mathbf P_i = 0$ and $\mathbf D_\perp = \mathbf D_\perp^*$ in
Eq.~\ref{cross-section})
\begin{equation}
\sigma=N^2+D_\perp^2.
\end{equation}
The second master equation, Eq.~\ref{polar}, yields for $\mathbf
P_i = 0$
\begin{equation}
\mathbf P_f={2N\mathbf D_\perp \over \sigma} = {2N\mathbf D_\perp
\over N^2+D_\perp^2}. \label{heusler}
\end{equation}
Hence, the diffracted beam from a single-crystal is completely
polarized if there is a Bragg reflection with $|\mathbf D_\perp| =
|N|$. Typical examples are the (111) reflection of Heusler $\rm
Cu_2MnAl$ ($d$-spacing $\rm 3.43$ \AA) and the (200) reflection of
the alloy $\rm Co_{0.92}Fe_{0.08}$ ($d$-spacing $\rm 1.76$ \AA).
Other single crystals like $\rm Fe_3O_4$ or $\rm Fe_3Si$ have also
been considered but are less used. All these crystals can be used
to produce polarized and monochromatic neutron beams and to
analyze the energy and polarization of neutron beams. Therefore,
single-crystal polarizers are used for single-crystal diffractometers and
in triple-axis
spectroscopy. Depending on the requirements on neutron energy and
resolution, different $d$-spacings must be considered. Recently,
the quality and the reflectivity of Heusler monochromators has
been improved considerably \cite{courtois} that will allow to use
these crystals at relatively short neutron wavelengths.

\subsection{Thin films}

Total reflection from magnetized thin films can be used to produce
polarized neutrons. The angle of total reflection for a
ferromagnetic film is given by
\begin{equation}
\theta^\pm_c=\lambda\sqrt{N(b\pm p)/\pi}, \label{critical}
\end{equation}
where $\lambda$ is the neutron wavelength. $N$ is the nuclear
density, and $b$ and $p$ the nuclear and magnetic scattering
lengths, respectively. Thus, by an appropriate choice of
materials, a polarized beam can be produced by total reflection.
For the special case $b = p$ all reflected neutrons are polarized.
Unfortunately, the reflection angles are only reasonably large for
cold neutrons: For example, Fe$_{50}$Co$_{48}$V$_2$ has $b \simeq
p$ and $\theta_c \simeq 0.4^0$ for $\lambda \simeq 4$ \AA\
\cite{schaerpf}.

The angles of reflection can be significantly improved by adding
artificial magnetic and non-magnetic layers that reflect neutrons
at small angles above $\theta_c$. Such artificial multi-layers
(supermirrors) have been produced by physical vapor deposition by
Mezei for the first time \cite{mezei}. Typical materials
combinations are Co/Ti, Fe/Si, and $\rm
Fe_{50}Co_{48}V_2/TiNi_{x}$ \cite{anderson,
krist,schebetov,hoghoj99,majkrzak}. The latter combination
exhibits a remanent magnetization and can therefore be used as a
spin selective device, where no spin flipper is necessary anymore
\cite{boni}. Recently, the number of layers has been increased
steadily thus leading to reflection angles for polarized neutrons
of the order of $0.3^0$ for $\lambda \simeq 1$ \AA. These modern
devices can now also be used as white beam polarizers for thermal
neutrons.

\subsection{Spin filters}

A major drawback of polarizing single-crystals and thin films is
their decreasing efficiency with increasing neutron energy, i.e.
short wavelength, and the small divergence that they accept (see
Table \ref{perf}). On the other hand they are maintenance free and
easy to use. Therefore, polarizing filters with broad-band
characteristics and minor restrictions on divergence are of
significant interest for neutron scattering in particular for
pulsed spallation sources and spectrometers with large
area-detectors.

\begin{table}[htb]
\caption{Performance and applications of various neutron
polarizers. The quoted
         values are only approximate. The notation is TAS: triple-axis spectrometer,
         DAX: double axis diffractometer, NSE: neutron-spin-echo, REF: reflectometer,
         TOF: time-of-flight, SANS: small angle neutron scattering. The quality factor
         is defined by $Q = TP^2$, where the transmission/reflection $T$ and
         polarization $P$ are taken from the literature.}
\vskip 2 mm \label{polarisers}
\begin{tabular}{c|ccccccc}
\hline \hline
 technique    &beam            &$E$-range (meV) &instruments    &$T$    &$P$ &$Q$ &ref.  \\
\hline
 Heusler      &fixed $\lambda$ &$    E < 80$    &TAS,DAX             &$0.62$ &0.95  &0.56 &\cite{courtois} \\
 super-mirror  &white           &$    E < 20$    &NSE, TAS, REF, TOF  &$0.9$  &0.95  &0.81 &\cite{hoghoj99}  \\
 $^3$He       &white           &$   E < 2000$   &DAX, TAS, TOF, SANS &$0.4$  &0.80  &0.26 &\cite{heil99} \\
 p targets    &white           &$ E \gg 2000$   &SANS                &$0.4$  &0.80  &0.26 &\cite{heil99} \\
 SmCo$_5$     &white           &$20 < E < 180$  &not implemented     &$0.3$  &0.75  &0.17 &\cite{mayers85} \\
\hline \hline \label{perf}
\end{tabular}
\end{table}

Whereas Heusler and supermirrors are well established but still
progressing techniques, the $^3$He spin filters using direct
optical pumping of meta-stable $^3$He \cite{colegrove63} have
improved during the last few years \cite{heil99}. The basic idea
behind the filter-technique is the polarization dependence of the
transmission that can be written in its most simple form as
\cite{williams}
\begin{equation}
T(\lambda) = \exp{(-\sigma_0 Nd)}\cosh (\sigma_pNd),
\end{equation}
where $\lambda$ is the neutron wavelength, $d$ the thickness of
the filter and $N$ the $\rm ^3He$
density. $\sigma_0$ and $\sigma_p$ are the spin-independent and
polarization dependent cross-sections of $\rm ^3He$, respectively.
If the filter is not perfectly polarized, a significant portion of
the correctly polarized neutrons will be absorbed. Because the
absorption increases with increasing $\lambda$, the thickness $d$
of the filter must be optimized for the wavelength band to be used
even if $\sigma_0$ is small.

$\rm ^3$He filters are now in regular use at the ILL on several
instruments. Due to wall relaxation of polarized He nuclei, the
polarization and transmission of the filters decreases with time
and they have to be exchanged almost daily. It is foreseen that $\rm ^3$He
filters will further improve and find applications in particular
at pulsed neutron sources and instruments with large area
detectors. The method of using spin exchange of $\rm ^3$He with
optically pumped Rb vapor \cite{bouchiat60} is progressing too and
may challenge the meta-stable type of pumping. One major advantage
of the latter technique is that the filter has not to be exchanged
during an experiment.

The development of SmCo$_5$ polarizing filters has been conducted
at ISIS. If the problems of depolarization of the neutrons within
the filter and of $\gamma$-heating in intense neutron beams can be
solved one may obtain a quality factor of $Q \simeq 0.25$
\cite{mayers85} that is lower than the maximum to be achieved for
a future $^3$He filter.

Polarized hydrogen can also be used as polarizing filter.
Moreover, the spin dependent interaction of the neutrons with the
protons can be used for contrast variation of hydrogen-containing
materials in small angle neutron scattering experiments
\cite{stuhrmann99}. We defer the interested reader to the
literature.

\section{Determination of Form Factors and Spin Densities}
\label{mff}

Polarized neutrons allow to measure magnetic densities with
improved accuracy as compared to standard diffraction methods. The
method presented below applies to magnetic structures described by
a propagation vector $\mathbf Q_0=0$. For a paramagnet, a
ferromagnetic component can be induced by applying an external
magnetic field. According to Eq.~\ref{cross-section} in the case
of mixed nuclear-magnetic Bragg reflections with real structure
factors, the intensity ratio $R$ of scattered neutrons polarized
by an external magnetic field along the $+z$ or $-z$ direction,
where $z$ is perpendicular to the scattering plane (that contains
$\mathbf Q$), is given by
\begin{equation}
R={{I^{+z}}\over{I^{-z}}}={N^2+2ND_\perp^z+D_\perp^{z2}\over
N^2-2ND_\perp^z+D_\perp^{z2}} = {(N+D_\perp^z)^2\over
(N-D_\perp^z)^2}.
\end{equation}
$D_\perp^z$ is the projection of the magnetic interaction vector
$\mathbf D_\perp$ along the z-axis. Determination of a
spin density with polarized neutrons consists of
measuring $R$ at many different Bragg reflections $(hkl)$. As the
crystal structure and hence the chemical structure factors $N$ are
presumably known ($N$ depends on the Miller indices), the method
provides usually directly the values of the magnetic structure
factors. For small magnetic amplitudes, polarized neutrons give an
enhanced sensitivity compared to unpolarized neutrons that yield
an intensity $I_{np}=N^2 + {2\over3}D_\perp^2$. The factor $2/3$
comes from the spherical averaging of $\bf D_\perp$ in the second
term of Eq.~\ref{cross-section} over all directions with respect
to $\bf Q$ for a non-magnetized isotropic sample.

Namely, considering a typical example with $D_\perp=0.1N$ yields
$I_{np} \simeq 1.01N^2$, while the contrast as measured with
polarized neutrons, $R=1.21N^2/0.81N^2=1.49$, is rather large.
Therefore polarized neutrons are particularly well suited for
measuring maps in compounds with small magnetic moments (example:
heavy fermion systems) or with seriously diluted magnetic moments
(example: molecular magnetic crystals).

To extract the magnetic moment density from the data, a classical
Fourier calculation is usually performed \cite{schweizer}. Because
\begin{equation}
\rho(\mathbf Q)=\int \int \int{m(\mathbf r)\exp(i\mathbf Q \cdot
\mathbf r)d^3\mathbf r},
\end{equation}
one can obtain the spin density $m(\mathbf r)$ in real
space by the inverse Fourier transform through the relation
\begin{equation}
m(\mathbf r)={1\over V}\sum_{\mathbf Q}{\rho(\mathbf
Q)\exp(-i\mathbf Q \cdot \mathbf r)}.
\end{equation}
As the cloud of unpaired electrons that are responsible for
magnetism is extended in real space, the magnetic form factor
decreases with increasing $\mathbf Q$ and equivalently with
increasing Bragg indices $(hkl)$. To obtain precise measurements,
data are to be taken up to large values of scattering vectors
$\mathbf Q$. To that end short-wavelength neutrons are to be
preferred, and instruments dedicated to such measurements provide
usually hot neutrons, like the 2-axis diffractometers D3 at the
ILL and 5C1 at the LLB. The layout of such an instrument is
presented in Fig. \ref{d3}. Until now, no polarized neutron
diffractometers have been built at a spallation source. In any
case, however, the data set is restricted to finite values of
$h,~k,~l,$ which  leads to oscillatory distortions in the spin
density maps $m(\mathbf r)$ due to finite size effects
\cite{schweizer}. Therefore, other reconstruction methods of the
spin density map have been developed. The two most often used
techniques are either based on information theory like the maximum
entropy method \cite{papoular} or on the multi-polar expansion of
the electronic density \cite{gillon}. The latter method models the
spin density by a set of parameters that have to be determined by
standard least-square fitting calculations.

\begin{figure}[h]
\vspace{0 mm} \caption
   {Schematic arrangement of a two-axis spectrometer used for the determination of
    magnetic densities. The monochromator produces a polarized beam with the neutron polarization
    perpendicular to the scattering plane.
    A small guide field prevents the neutron beam to
    depolarize.
    A \textit{spin-flipper} allows to reverse the neutron polarization by $180^0$ and hence to
    measure the flipping ratio $R$.
    A magnetic field saturates the magnetic moments of the sample along the neutron polarization.}
\label{d3}
\end{figure}

\subsection{Magnetic Form Factors}
Ever since the experiments of Shull and coworkers in the 1960's in
3d ferromagnets \cite{shull}, the main motivation to measure spin
densities has been to gain a better insight in electronic
distributions in solid-state materials. Since that time,
measurements have been extended to paramagnetic metals and to the $4f$-electrons in
rare-earth compounds \cite{moon2,boucherle}.

In $3d$-ferromagnets whereas the atomic form factor can be very
precisely reproduced from spin-polarized Hartree-Fock
calculations \cite{watson}, there is a strong indication from form factor measurements
of a
negative spin polarisation between the atomic sites in both Fe and
Co. Also, the magnetic moment density-map in Ni
\cite{mook} and Pd \cite{cable} shows an aspherical d-electron
distribution plus an orbital contribution. In order to allow for
these effects, the form factor is usually written as
\begin{eqnarray}
f(\mathbf Q) &= &{2\over g} (1+\alpha)\biggl[\langle j_0 \rangle
+({5\over2}\gamma -1)A_{hkl}\langle j_4 \rangle\biggr]
 \nonumber \\
 & & +\biggl[{(g-2)\over g} f_{orb}-({2\over g})\alpha\delta(\mathbf Q)\biggr],
\end{eqnarray}
where $g$ is the Land$\acute{e}$ factor; $\alpha$ is a parameter describing the
fraction of negative spin polarization; $\gamma$ is the percentage of electrons in
$E_g$ orbitals which takes into account the orbital contribution \cite{mook}. $A_{hkl}$ is a
geometrical factor; $\langle j_0 \rangle$ and $\langle j_4 \rangle$ represent the spherical
and aspherical part of the form factor, respectively. A comparison of the calculated
and measured form factors for Ni yielding a uniform negative contribution equal to
$-0.0091 \mu_B/\AA^3$ is shown in Fig.~\ref{niff}. The spin magnetic moment
per Ni-atom is $\mu_{spin}=0.656 \mu_SB$ and the orbital contribution $\mu_{orbital}=0.055 \mu_B$.
\begin{figure}[h]
\vspace{0 mm} \caption
   {Magnetic moment distribution of Ni in the [100] plane (taken from Ref. \cite{mook}).}
\label{niff}
\end{figure}

On the contrary to metallic compounds with $d$-electrons,
$f$-electrons are well localised around the nuclei which allows to
perform atomic calculations to obtain the spin distribution. As
the orbital moment is usually different from zero, there is an
significant contribution of the orbitals to the magnetic density.
Also, the spin-orbit coupling is important which results in a
mixing of the atomic wave functions. The form factors for the
atoms of the rare-earth and actinide series have been calculated
by Desclaux and Freeman \cite{desclaux} using the relativistic
Dirac-Fock theory. It was shown that the atomic form factor can be
expressed as
\begin{equation}
f(\mathbf Q)=\langle j_0 \rangle + c_2 \langle j_2 \rangle + c_4 \langle j_4 \rangle
+ c_6 \langle j_6 \rangle,
\end{equation}
where
\begin{equation}
\langle j_l(\mathbf Q)\rangle=\int_0^\infty{U^2(r)j_l(Qr)4\pi^2dr}.
\end{equation}
$U(r)$ is the radial wave-function for the unpaired electrons in the atom, and $j_l(Qr)$ the
Bessel function of $l$th-order. The coefficients $c_i$ are tabulated in e.g.
Ref. \cite{tableinter}.

Among the rare-earth elements, Samarium represents a particular
case as the orbital and spin contributions to the magnetisation
almost cancel out so that the magnetic density map contains both
positive and negative regions. This leads to a form factor which
has a maximum  located at a position different from $Q=0$. The Sm
form factor measured in $\rm SmCo_5$ is shown in Fig. \ref{smff}.
An interesting effect is, that as the first excited crystal-field
states are located at relative low energies, they become populated
when the temperature approaches 300 K. Consequently, the magnetic
moment of Sm is only $\mu\sim 0.04\mu_B$ at room temperature and
increases to $\mu \sim 0.38 \mu_B$ for $T=4.2$ K \cite{givord}.
For all temperatures, however, the form factor of Sm has a strong
orbital character.
\begin{figure}[h]
\vspace{0 mm} \caption
   {a) Experimental form factor for Sm at $T=4.2$ K. The line corresponds to a calculation
including crystal field, exchange and spin-orbit effects. b) Same
for $T=300$ K (taken from Ref. \cite{givord}).} \label{smff}
\end{figure}

Finally, we should point out that a study of the spatial
distribution and temperature dependence of the spin density allows
to probe the spin susceptibility $\chi (\mathbf Q, 0)$. As such
the method can  been used to investigate the nature of the
electrons \textit{e.g.} in superconductors. For example, in $\rm
V_3Si$ \cite{shull2} it was shown that the spin susceptibility of
the $\rm V$ electrons disappears upon entering the superconducting
phase which is an  indication of  spin-pairing. On the other hand,
no similar effect could be observed in the new heavy-fermion
superconductors $\rm UPt_3$, $\rm UBe_{13}$ and $\rm CeCu_2Si_2$.
For the latter compounds, the spin susceptibility is temperature
independent in the superconducting phase \cite{stassis}. These
results are of particular importance as they impose severe
restrictions on the possible pairing mechanisms that can give rise
to the electron pairing in these unconventional heavy-fermion
superconductors.

\subsection{Magnetisation distribution in molecular magnets}
Molecular magnetism is a fast growing field in material science with potential important
technological applications in electronic devices. By building blocks of
molecules which contain magnetic centres, magnetic interactions can be tuned and the aim
is to synthesize organic compounds which exhibit magnetic ordering  at room temperature.
To that end, as the number of combinations
offered by organic chemistry is almost infinite, it is essential that the mechanism
of magnetic couplings originating from $2p$ electrons to be well understood \cite{kahn}.

In contrast to ionic systems where the electrons which carry
magnetism are well localised around the nuclei, the magnetic
density of organic compounds is distributed over all molecules due
to covalency effects. The effect of delocalisation is even more
pronounced when there is no magnetic ion in the molecule and
magnetism is due to $2p$ electrons only \cite{schweizer97}.
Polarised neutron diffraction yields directly the distribution of
electrons responsible for magnetism in organic materials which in
turn can be directly compared to theoretical calculations for the
electronic wave-functions and the chemical bonds
\cite{ressouche99}. Spin density in molecular compounds can also
be used to trace exchange pathways through the molecules, like
when spin polarisation is found on atoms which in principle are
non-magnetic. This is for example the case in the free radical
nitronyl nitroxides NitPy(C$\equiv$C-H).
 NitPy(C$\equiv$C-H) builds zig-zag chains linked by C$\equiv$C-H$\cdots$O pieces
where the hydrogen bridge two molecules. Significant spin
population is found at the hydrogen positions ($\mu\sim0.04\mu_B$) which
indicates that the hydrogen bond is involved in the ferromagnetic exchange interactions between
the molecules \cite{ressouche97}.

\begin{figure}[h]
\vspace{0 mm} \caption
   {Spin density projection in NitPy(C$\equiv$C-H). The contour step for the pyridine
cycle is $\rm 0.008 \mu_B/\AA^2$ whereas for the Nit cycle a step is equal to $\rm 0.04 \mu_B/\AA^2$
   (taken from Ref. \cite{ressouche97}).}
\label{NitPy}
\end{figure}

A typical example of \textit{spin delocalisation} is found in the compound
$\rm MnCu(pba) (H_2O)_3\cdot2H_2O$, with pba=1,3-propylenebis(oxamato).
The $\rm Mn^{2+}$ and $\rm Cu^{2+}$ ions are connected by oxamato bridges
and build ferrimagnetic chains. The $\rm Mn^{2+}$ ions carry a spin $\rm S_{Mn}={5\over 2}$
and the $\rm Cu^{2+}$ have an effective spin $\rm S_{Cu}={1\over 2}$.
Therefore we have the situation where two magnetic metals are linked by organic species.
The magnetisation, as obtained from polarised neutron diffraction \cite{baron, baron2}, shows a
positive spin population (i.e. the induced magnetisation is aligned along the
applied magnetic field) for the Mn spins
whereas it is negative for the copper magnetic moments. This reveals the
antiferromagnetic nature of the intra-chain coupling.
Interestingly, an important contribution to the spin density map is found on
the neighboring oxygen and nitrogen ions and on the two central carbon atoms. Summing up the
positive and negative spin polarisations individually, one obtains $\rm 5.1 \mu_B$
and $\rm -1.0 \mu_B$, respectively, which shows that the metallic ions have distributed their
spin densities on the molecule. The magnetic moments distribution for
$\rm MnCu(pba) (H_2O)_3\cdot2H_2O$ is shown in Fig. \ref{mncu}.

\begin{figure}[h]
\vspace{0 mm} \caption
   {a) Experimental spin density map of $\rm MnCu(pba) (H_2O)_3\cdot2H_2O$. The contour step is
   $\rm 0.005 \mu_B/\AA^2$. The continuous line represents the
   positive spin distribution while the dotted line describes the negative magnetisation. 
    b) Calculated spin density map for an isolated molecule with the DMol$^3$ method 
       (after Ref.~\cite{baron2}). The lowest contour is at $\rm 0.005\mu_B/\AA^2$.}
\label{mncu}
\end{figure}
Fig.~\ref{mncu} also shows the theoretical spin density for the 'CuMn' molecule
projected onto the oxamid mean plane. The theory is based on the local-spin 
density-functional principles of Perdew and Wang \cite{perdew}. 
Calculations for the cation ($2+$) in vacuum are done with the DMol$^3$ method \cite{delley}.
On comparing with experiment it is clear that there is a disagreement
with theory on the sign of the spin density
at the bridging carbon atoms. On should remember however, that the theory
applies to an isolated cation in vacuum. Calculations for smaller than the formal charge 
reverse the spin density at the bridging carbons.
The crystal environment may also change areas with small spin density.

\subsection{Spin susceptibility in the high-T$_c$ superconductor $\rm YBa_2Cu_3O_{7-x}$}
The discovery of the high-T$_c$ superconductor $\rm
La_{2-x}Ba_xCuO_4$ with $\rm T_c=35K$ by Bednorz and M\"uller
\cite{bednorz} in 1986 has been at the origin of an enormous
amount of work to understand the electronic (charge and spin)
correlations in these materials. Following the discovery of the
$\rm La_{2-x}Ba_xCuO_4$ compound, other materials exhibiting
similar or higher transition temperatures for superconductivity
have been synthesized, like $\rm La_{2-x}Sr_xCuO_4$,$\rm
Nd_{2-x}Ce_xCuO_4$, $\rm YBa_2Cu_3O_{7-x}$, and others. All these
materials share common features of their crystallographic
structure. They possess $\rm CuO_2$-layers well separated from
each other, so that they can be considered as
quasi-two-dimensional materials. In this class of materials,
superconductivity is achieved by carefully tuning the amount $x$
of $\rm Sr$, $\rm Ce$ or $\rm O$ which results in doping the $\rm
CuO_2$-layers with charge carriers. The important feature is that
the cuprate materials are \textit{either} antiferromagnets and
insulators \textit{or} paramagnetic metals and superconductors
below a critical temperature $\rm T_c$. Important
antiferromagnetic correlations and fluctuations persist in the
superconducting phase. The role played by these fluctuations in
the formation of the superconducting state is still the subject of
an intense debate. A central piece of the physics of the
high-T$_c$ superconductors is the understanding of the charge and
spin states in the $\rm CuO_2$-layers as a function of doping both
below and above the transition temperature $\rm T_c$.

The intensity of the scattered neutrons can be directly related to
the spin susceptibility at $\mathbf{Q}=0$ and $\rm \omega=0$
through the dissipation-fluctuation theorem. As the signal is
particularly small in the high-T$_c$ compounds, use of
polarisation analysis is required to enhance the contrast and to
isolate the weak magnetic contribution. The results obtained for
the temperature variation of the local susceptibility at the
copper sites in the $\rm CuO_2$ layers is shown in Fig. \ref{cu2}.
A particularity of the temperature dependence of the signal is the
appearance of the so-called \textit{spin pseudo-gap} in
under-doped $\rm YBa_2CuO_{6.52}$ for which the spin
susceptibility drops above T$_c$. On the other hand, the local
spin susceptibility in optimal doped samples decreases only upon
cooling below the superconducting temperature \cite{rossat}.

\begin{figure}[h]
\vspace{0 mm} \caption
   {Local spin susceptibility as a function of temperature on copper sites
    in the $\rm CuO_2$ layers in the high-T$_c$
    superconductors $\rm YBa_2Cu_3O_{7-x}$ (taken from Ref. \cite{rossat}).}
\label{cu2}
\end{figure}

\section{Spherical Neutron Polarimetry}

Spherical Neutron Polarimetry (SNP) has recently been developed
and successfully tested at the ILL \cite{brown} as an alternative
way of measuring magnetic structures. Moreover, this method allows
to determine form factors and spin densities in antiferromagnets
for which very few data is available. The classical technique
discussed in chapter \ref{mff} cannot be applied in
antiferromagnets with propagation vector $\rm \mathbf Q_0=0$
when the magnetic and nuclear structure factors are in phase
quadrature. For such cases, like $\rm Cr_2O_3$ or even hematite,
the neutron cross-section $\sigma$ is
polarisation independent  \cite{forsyth}
\begin{equation}
\mathbf P_f\sigma=\mathbf P_0(1-\gamma^2)+2\gamma^2\mathbf{\hat Q}(\mathbf P_0\cdot
\mathbf{\hat Q})+2\gamma(\mathbf P_0 \times \mathbf{\hat Q}),
\end{equation}
with $\sigma=1+\gamma^2$ and $\gamma \mathbf{\hat Q}=\Im \mathbf
D_{\perp}(\mathbf Q)/N$. SNP gives access to the complete set of
independent correlation functions involved in the nuclear-magnetic
scattering process by a direct measurement of the three components
of the polarization vector $\mathbf P_f$ of the scattered
neutrons. Eq. \ref{polar} shows that if the polarization $\mathbf
P_i$ of the incoming neutron beam is fixed, a measurement of
$\mathbf P_f$ allows in most cases an unambiguous determination of
the direction of the magnetic interaction vector $\rm \mathbf
D_\perp$. This is an alternative way of determining magnetic
structure factors to the standard diffraction method that relies
on a precise measurement of neutron intensities. Measuring
intensities is the same as measuring $\rm \mathbf
D_\perp\cdot\mathbf D_\perp^*$ which leads to a loss of phase
factors and often magnetic structures cannot be unambiguously
resolved by unpolarized neutron diffraction. SNP has been
successfully applied  in problems involving complex magnetic
structures, like spiral structures, systems with magnetic domains
and small magnetic moments, frustrated antiferromagnets, etc. It
has to be pointed out that the method is sensitive to the
direction of the magnetic interaction vector only and not to its
magnitude. In this case, finding the value of magnetic moments
requires, as usual, the comparison of magnetic and nuclear
cross-sections. In contrast to standard single-crystal
diffraction, it is insensitive to secondary extinction and allows
magnetic structure determination of samples even in the presence
of magnetic domains. Namely, magnetic domains depolarize the
neutron beam according to their respective population. In other
words, the domain population is obtained by measuring the
amplitude of the neutron polarization vector ${\rm \mathbf P}_f$.
In order to perform spherical neutron polarimetry a non-isotropic
domain distribution is usually necessary. It can be induced for
example by the application of uniaxial pressure.

Spherical neutron polarimetry out-performs standard polarized
neutron scattering as it allows to measure both the longitudinal
and transverse components of ${\rm \mathbf P}_f$. Namely, if a
magnetic field is applied to the sample, as it is the case for the
longitudinal polarimetry explained in section \ref{mff}, only the
component of the polarization longitudinal to the field can be
measured. The transverse components depolarize rapidly and are
lost \cite{newton}. This is the case in antiferromagnets with
mixed nuclear-magnetic Bragg reflections, where most information
is contained in the transverse components through the
nuclear-magnetic interference term.

\subsection{Realization of a zero-field chamber: Cryopad}

Following the introduction in the previous section it is clear
that the transverse components of the polarization can only be
measured if the sample is placed in a zero-field chamber. Such a
device (called Cryopad) has been constructed at the ILL
(Fig.~\ref{cryopad}) \cite{tasset}. It consists of two cylindric
Meissner shields in the superconducting state. The diameter of the
inner shield is large enough to accommodate a cryostat and/or
other devices to define the sample environment. Cryopad is
centered on the sample table and its orientation is fixed with
respect to the wave-vector ${\rm \mathbf k}_i$ of the incident
neutrons. Therefore, the sample can be oriented independently form
Cryopad in order to access various Bragg-reflections.

\begin{figure}[h]
\vspace{0 mm} \caption
   {Schematic zero-field chamber \textit{Cryopad II} used for spherical polarimetry at ILL
   (Taken from Ref. \cite{tasset}).}
\label{cryopad}
\end{figure}

The components of the polarization of the incident and scattered
neutrons are defined independently by means of two rotating
solenoids (called \textit{nutators}) that are placed in the
incident and scattered neutron beam. They act as guide-fields in
order to orient the neutron polarization vectors $\mathbf
P_\alpha$ ($\alpha =$ incident, final) in the plane parallel to
the Meissner shields (transverse to $\mathbf k_\alpha$). Two spin
turning coils between the Meissner shields apply a horizontal
field transverse to $\rm \mathbf k_\alpha$ and allow the
definition of the component of $\mathbf P_\alpha$ along $\rm
\mathbf k_\alpha$. The modulus of $\mathbf P_f$ is determined by
measuring the flipping ratio of the scattered neutrons simply by
reversing the field of the nutator after the sample. This can be
accomplished quickly by reversing the current in the solenoid. The
combined use of the nutators and of the spin turning coils in
Cryopad allows the analysis of $\mathbf P_f$ for any direction of
$\mathbf P_i$. In practice, the information is obtained by
measuring the three orthogonal components of the final
polarization $P_f^\alpha$ ($\alpha = x,y,z$) for the three
orthogonal directions of the initial polarization $\mathbf P_i$.
The direction $x$ is defined as being along the scattering vector
$\mathbf Q$, $z$ is chosen perpendicular to the scattering plane
and $y$ is the last orthogonal direction. For example, if the
initial polarization is chosen along the $z$-axis, the component
of the final polarization along the $x$-direction is given by
$P_x=(n_{x+} - n_{x-})/(n_{x+} + n_{x-})$, where $n_{x+}$ and
$n_{x-}$ are the number of neutrons with spins along or
anti-parallel to the $x$-axis, respectively \cite{tasset}.
Measuring the nine flipping ratios is sufficient to determine the
required information about the orientation of the magnetic
interaction vector $\mathbf D_\perp (\mathbf Q)$ and to obtain the
value of the ratio between magnetic and nuclear amplitudes
$|\mathbf D_\perp (\mathbf Q)|/N$. Geometrical relationships
between the direction of the final neutron polarization relative
to $\mathbf P_i$ have been derived by Nunez \textit{et al.}
\cite{nunez} and are of great help to determine the direction of
$\mathbf D_\perp (\mathbf Q)$.

\subsection{Example: UPtGe}

Heavy-fermion materials are characterized by a very large linear
coefficient of the specific heat and a greatly enhanced Pauli
susceptibility, corresponding to effective masses of the
quasi-particles of about two orders of magnitude larger than the
free electron. Heavy fermions are therefore ideal systems to study
strong electron correlations and the number of compounds showing
heavy-fermion behavior is large. The ground-state of these systems
varies from metallic ($\rm CeCu_6$, $\rm CeIn_3$) to insulating
($\rm Ce_3Bi_4Pt_3$) and from antiferromagnetic ($\rm U_2Zn_{17}$,
$\rm UPd_2Al_3$, $\rm UNi_2Al_3$) to superconducting ($\rm UPt_3$,
$\rm UPd_2Al_3$, $\rm UNi_2Al_3$, $\rm CeCu_2Si_2$). In most of
these systems the magnetic ground state is determined by the
competition between the Kondo-effect that tries to screen the
magnetic moment and the RKKY interaction that tends to stabilize a
ground-state with long-range magnetic order.

Non-collinear magnetic structures in compounds with localized
spin-densities can be explained on the basis of the Heisenberg
model to originate from competition between exchange forces. In
systems with \textit{5f}-electrons, like $\rm U_3P_4$ or $\rm
U_2Pd_2Sn$ it has been shown by calculations based  on the
local-spin density functional theory that a non-collinear
arrangement of the magnetic moments is the consequence of strong
spin-orbit coupling \cite{sandratskii}. However, this theory does
not favor the helicoidal-type of magnetic structures found in e.g.
$\rm UNi_2Al_3$ and $\rm UPtGe$.

As an example of a magnetic structure determination with the help
of spherical neutron polarimetry we show results obtained from the
ternary compound $\rm UPtGe$ that orders below $T_N
\sim 50$ K with a propagation vector $\mathbf Q_0=(0.554,0,0)$
\cite{robinson}. Measurements in single crystals using unpolarized
neutrons could not decide between an amplitude-modulated
spin-density wave and a cycloid with unequal magnetic moments
along the $a$- and $c$-axis (ellipticity), respectively, yielding
for the two models similar agreement factors between observed and
calculated structure factors \cite{robinson,szytula}.

It is seen from Eq.~\ref{polar} that the direction as well as the
amplitude of the polarization of the scattered neutrons depend
upon the value of the chiral term. In particular, the chiral
component disappears, when $\mathbf D_\perp(\mathbf Q_0)$ is
parallel to $\mathbf D_\perp^*(\mathbf Q_0)$, as it is the case for
an amplitude modulated wave and the polarization of the neutron
precesses by $\rm 180^\circ$ around the scattering vector
\cite{nunez}.
In contrast, the chiral term gives a contribution to $\mathbf P_f$
if the magnetic structure is a cycloid. Hence, the direction of
$\mathbf P_f$ depends on the scattering geometry.
Spherical neutron polarimetry therefore allows to distinguish
between an amplitude modulated spin-density wave and a helix
\cite{paixao}. For $\rm UPtGe$, the directions of polarization of
the diffracted neutrons for directions of $\mathbf P_i$
perpendicular to the scattering plane ($z$), along the scattering
vector ($x$), and a third direction in the scattering plane but
perpendicular to $x$  are summarized in table \ref{table1}. The
results of spherical neutron polarimetry unambiguously show that
the magnetic structure of $\rm UPtGe$ is a cycloid with an axis
ratio $\rm \sim 1.24$ \cite{mannix}, as shown in Fig. \ref{uptge}.

\begin{figure}[h]
\vspace{0 mm} \caption
   {Magnetic structure of $\rm UPtGe$ as determined by spherical neutron polarimetry using
    the Cryopad device (after \cite{mannix}).}
\label{uptge}
\end{figure}

\begin{table}[t]
\caption{Spherical Neutron Polarimetry data obtained with Cryopad
in UPtGe (after Ref. \cite{mannix}). $P_i$, $P_f$ and $P_{calc}$
are the incident, scattered and calculated neutron polarizations.
The polarization axis are defined as: $x$ is parallel to the
scattering vector $\bf Q$; $y$ is perpendicular to $\bf Q$ and in
the scattering plane; $z$ is vertical.} \label{table1}
\begin{center}
\begin{tabular}{l|ccc|ccc|ccc}
\hline {}                               &{}      &$P_i$   &{} &{}
&$P_f$    &{}                &{}      &$P_{calc}$ &{}
\\
\hline
 hkl                             &x      &y      &z            &x       &y      &z                 &x        &y      &z           \\
\hline \hline $\rm 0^+00$                      &0      &0
&0.9          &0.01    &-0.05  &0.91               &0       &0
&0.90           \\
                                 &0      &0.9    &0            &0.08    &-0.91  &-0.04              &0       &-0.90  &0           \\
$\rm 0^+20$                      &0      &0      &0.9
&0.94    &-0.11  &-0.12              &0.97    &0      &-0.11
\\
                                 &0      &0.9    &0            &0.94    &-0.02  &0.03               &0.97    &0.11   &0           \\
$\rm 2^-20$                      &0      &0      &0.9
&-0.91   &0      &0.16               &-0.95   &0      &0.20
\\
                                 &0      &0.9    &0            &-0.91   &-0.10  &-0.01              &-0.95   &-0.2   &0           \\
$\rm 0^-20$                      &0      &0      &0.9
&-0.93   &0.05   &-0.14              &-0.97   &0      &-0.11
\\
                                 &0      &0.9    &0            &-0.92   &0.13   &-0.02              &-0.97   &0.11   &0           \\
                                 &0      &0      &-0.9         &-0.92   &0.14   &0.09               &-0.97   &0      &0.11           \\
                                 &0      &-0.9   &0            &-0.93   &0      &-0.04              &-0.97   &-0.11  &0           \\
\hline
\end{tabular}
\end{center}
\end{table}

\section{Inelastic Neutron Scattering with Polarized Neutrons}

In analogy to the neutron cross-section derived by Blume \cite{blume} for the elastic case,
there are three contributions to the inelastic cross section:
\begin{itemize}
\item{a pure nuclear one that gives rise to phonon scattering},
\item{a pure magnetic one when neutrons are scattered \textit{e.g.} by spin waves,}
\item{and a magnetic-nuclear interference term that is present only in
      special cases, as \textit{e.g.} when the spin-lattice interaction in a
      ferromagnet cannot be neglected.}
\end{itemize}
The inelastic neutron cross-section and its relationship to the neutron polarisation have been
derived by many authors (see \textit{e.g.} \cite{lovesey, squires}).
In the following we will reproduce the calculations of Maleyev \cite{maleyev99} which
expresses the time-dependent scattering amplitudes in terms of the Van Hove correlation
function
\begin{equation}
\pi S_{AB}(\omega)={1\over{1-\exp (-\omega/T)}}<A, B>''_\omega,
\end{equation}
where $<A, B>''_\omega$ is the absorptive part of the generalised retarded susceptibility
\begin{equation}
<A, B>''_\omega=\pi(1-\exp (-\omega/T))Z^{-1}\sum_{a,b}\exp(-{E_a\over T})
A_{ab}B_{ba}\delta(\omega +E_{ab}).
\end{equation}
$A$ and $B$ are operators like $N({\bf Q})$ or ${\bf D}_\perp(\bf
Q)$ from Eq.~\ref{cross-section}; $Z^{-1}$ is the partition
function and $E_{a,b}$ are the energies between eigenstates of the
system. Expressed in such terms, the inelastic neutron
cross-section is given by
\begin{equation}
\sigma=\sigma_n+\sigma_m+\sigma_{nm},
\end{equation}
with
\begin{eqnarray}
% nuclear cross-section
\sigma_n &= &{{k_f}\over{ k_i}}{1\over \pi}{1\over{1-\exp
(-\omega/T)}}<N(-\mathbf{Q}), N(\mathbf{Q})>''_{\omega} \nonumber
\\ 
% magnetic cross-section
\sigma_m &= &{{k_f}\over{ k_i}}{1\over \pi}{1\over{1-\exp
(-\omega/T)}}\sum_{\alpha\beta\gamma}<D^\alpha_{\perp}(-\mathbf{Q}),
D^\beta_{\perp}(\mathbf{Q})>''_{\omega}
         (\delta_{\alpha\beta}+i\epsilon_{\alpha\beta\gamma}P_{i\gamma}) \nonumber \\
%nuclear-magnetic interference term
\sigma_{nm} &= &{{k_f}\over{ k_i}}{1\over \pi}{1\over{1-\exp
(-\omega/T)}}  (<N(-\mathbf Q), \mathbf D_{\perp}(\mathbf
Q)>''_\omega + <\mathbf D_{\perp}(-\mathbf Q), N(\mathbf
Q)>''_\omega) \cdot \mathbf{P}_i.
\end{eqnarray}
$\{\alpha, \beta, \gamma\}=\{x,y,z\}$ in Cartesian coordinates.
The polarisation vector $\mathbf P_{f}$ is accordingly given
\begin{itemize}
\item{for nuclear scattering by $\mathbf{P}_f\sigma_n=\mathbf{P}_i\sigma_n$,}
\item{for pure magnetic scattering by
\begin{eqnarray}
P_{f\alpha}\sigma_m &=
&\sum_{\beta}P_{i\beta}(<D_{\perp}^\alpha(-\mathbf Q),
D_{\perp}^\beta(\mathbf Q)>''_\omega
                    +<D_{\perp}^\beta(-\mathbf Q), D_{\perp}^\alpha(\mathbf Q)>''_\omega) \nonumber \\
                   & &-\sum_{\beta\gamma}\delta_{\alpha\beta}<D_{\perp}^\gamma(-\mathbf Q),
            D_{\perp}^\gamma(\mathbf Q)>''_\omega
                        -i\sum_{\beta\gamma}\epsilon_{\alpha\beta\gamma}
<D_{\perp}^\beta(-\mathbf Q),D_{\perp}^\gamma(\mathbf Q)>''_\omega, \nonumber
\end{eqnarray}
}
\item{for magnetic-nuclear interference scattering by
\begin{eqnarray}
\mathbf{P}_f\sigma_{nm}& = &<N(-\mathbf Q), \mathbf
     D_{\perp}(\mathbf Q)>''_{\omega} + <\mathbf D_{\perp}(-\mathbf Q), N(\mathbf Q)>''_{\omega} \nonumber \\
      & & +i[<N(-\mathbf Q), \mathbf D_{\perp}(\mathbf
      Q)>''_{\omega} - <\mathbf D_{\perp}(-\mathbf Q), N(\mathbf Q)>''_{\omega}]\times \mathbf{P}_i. \nonumber
\end{eqnarray}
 }
\end{itemize}
It turns out from these equations that whereas phonons do not
change the neutron polarisation, scattering  by spin waves do. As
we will see below this feature is very useful to separate and
identify the different magnetic modes in ferro- and
antiferromagnets. For the particular cases of simple ferromagnets
and of two-sublattices collinear antiferromagnets, explicit
expressions for the polarisation dependence of the inelastic
neutron cross-section can be found in the classic paper of Izyumov
and Maleev \cite{izyumov}. The magnetic-nuclear interference term
gives rise in particular to the so-called
\textit{magneto-vibrational} scattering and is also important for
a proper understanding of \textit{magneto-elastic} scattering.
Their origins are due to the fact that the cloud of electrons that
carries magnetism follows the nuclei when they oscillate around
their equilibrium position and that the magnetic moment is
modulated by the lattice vibrations, respectively. The
magneto-vibrational scattering is inelastic in the nuclear system
but elastic in the magnetic one. It occurs at the same positions
in reciprocal space as the phonons but with a polarisation
dependence. It has been exploited \textit{e.g.} to measure the
magnetic form factor through the polarisation dependence of
phonons at general $\mathbf{Q}$ positions in Fe \cite{steinsvoll}.
\subsection{Longitudinal neutron polarimetry}

The first spectrometer that allowed the analysis of the scattered
neutrons was built by Moon, Riste and Koehler by replacing
monochromator and analyzer of a triple-axis spectrometer by
ferromagnetic crystals that were saturated in a magnetic field
(Fig.~\ref{tasp}) \cite{moon1}. In contrast to spherical neutron
polarimetry the polarization is maintained by means of guide
fields throughout the instrument. As a consequence, only the
component of the neutron polarization parallel or anti-parallel to
the field direction can be measured whereas the transverse
components depolarize and are lost. In exactly the same way than
the magnetic-nuclear interference term cannot be measured by
standard diffraction, the same effect happens for the inelastic
counterpart. The magnetic-nuclear interference contribution leads
to a rotation of the initial polarisation which averages out if a
magnetic field is applied. Such a contribution is only accessible
if the sample is enclosed in a zero-field chamber, like the
Cryopad device.
 The standard triple-axis instrument with polarisation analysis developed by Moon et al. allows
to measure the energy, momentum and spin dependence of cross
sections with the restriction that only longitudinal polarimetry
can be performed. In practice, two basic
scattering geometries are commonly used: Namely the spin-flip and
non spin-flip cross sections are measured with either the
polarization of the neutrons parallel or perpendicular to the
scattering vector $\mathbf Q$.

\begin{figure}[h]
\vspace{0 mm} \caption
   {Schematic arrangement of the three axis spectrometer for polarized
   neutrons used by Moon et al. \protect{\cite{moon1}}, at Oak Ridge National
   Laboratory.}
\label{tasp}
\end{figure}

%                 \mbox{\boldmath$\sigma$}
%
An immediate application of longitudinal polarimetry is that for
$\mathbf P_i \parallel \mathbf Q$ all magnetic scattering involves
processes in which the spin of the neutrons is flipped because
$D_{\perp z} = 0$. We point out again (see Eq.~\ref{pola}) that
this is a general rule valid for elastic and inelastic as well as
coherent and incoherent scattering. In contrast, if the scattering
geometry is chosen such that $\mathbf P_i \perp \mathbf Q$ and
$\mathbf M \parallel \mathbf P_i$ then the elastic magnetic
scattering ($D_x=D_y=0$) is non spin-flip, whereas the inelastic
scattering is spin-flip ($\delta\mathbf D_y$, transverse
excitations) and non spin-flip ($\delta\mathbf D_z$, longitudinal
excitations), respectively.

Polarized neutrons in inelastic neutron scattering are often used
to separate the magnetic from the nuclear scattering or to
distinguish  magnetic fluctuations perpendicular and transverse to
the magnetization or scattering vector. When scattering by phonons
dominates a neutron spectrum an unambiguous determination of the
magnetic contribution to the neutron cross-section can be
accomplished by measuring the scattering with $\mathbf P_i
\parallel \mathbf Q$. A typical example for UFe$_2$ is shown in
Fig.~\ref{ufe2}, where the linear dispersion curve of the acoustic
phonons can be distinguished from the quadratic dispersion curve
of the spin-wave branch in $\rm UFe_2$ at low $T$ in the
cross-over regime \cite{paolasini}.

\begin{figure}[h]
\vspace{0 mm} \caption
   {Constant-energy inelastic scan in $\rm UFe_2$ using polarization analysis
    showing that spin waves occur
    in the spin-flip channel (black symbols), while scattering by phonons is non spin-flip (open
    symbols) (taken from Ref. \cite {paolasini}).
    The line is simply to guide the eye. See text for details.}
\label{ufe2}
\end{figure}

Longitudinal polarimetry is not only an important
method for measuring magnetic and nuclear cross sections
unambiguously, it is also very powerful in separating self and
collective dynamics in materials that contain strong incoherent
scatterers like hydrogen in biological materials and polymers
\cite{cook} (see chapter~\ref{cohincoh}).

\subsection{The XYZ method}
\label{xyzxyz}

The longitudinal polarimetry method
can be generalized to polarization analysis along the three
Cartesian directions, the
so called XYZ method (that is however to be distinguished from spherical
polarimetry). This method allows to apply the technique of
polarization analysis to time-of-flight spectrometers with
multi-detectors. With the coordinate system shown in Fig. \ref{xyz
geometry}, the non-spin flip (NSF) and spin-flip (SF)
cross-sections are given by \cite{scharpf}
\begin{eqnarray}
 \sigma^x_{NSF} &=& {1\over 3}\sigma_{NS}+{1\over 2}\sigma_M \sin^2\alpha +\sigma_N \\
 \sigma^y_{NSF} &=& {1\over 3}\sigma_{NS}+{1\over 2}\sigma_M \cos^2\alpha +\sigma_N \\
 \sigma^z_{NSF} &=& {1\over 3}\sigma_{NS}+{1\over 2}\sigma_M +\sigma_N \\
 \sigma^x_{SF}  &=& {2\over 3}\sigma_{NS}+{1\over 2}\sigma_M(1+\cos^2\alpha) \\
 \sigma^y_{SF}  &=& {2\over 3}\sigma_{NS}+{1\over 2}\sigma_M(1+\sin^2\alpha) \\
 \sigma^z_{SF}  &=& {2\over 3}\sigma_{NS}+{1\over 2}\sigma_M.
\end{eqnarray}
Here, $\sigma_M$ is the magnetic, $\sigma_{NS}$ is the nuclear
spin incoherent and $\sigma_N$ is the coherent plus isotopic
incoherent nuclear scattering cross-section. The magnetic
scattering can be isolated by combining these equations in the
following way (independent of the angle $\alpha$)
\begin{eqnarray}
{1\over 2}\sigma_M &=&
2\sigma^z_{NSF}-\sigma^y_{NSF}-\sigma^x_{NSF} \\
                   &=& \sigma^x_{SF}+\sigma^y_{SF}-2\sigma^z_{SF}.
\end{eqnarray}
The XYZ-method has been applied successfully in determining the
dynamical magnetic response in metals with small magnetic moments
like $\rm V_2O_3$ \cite{ziebeck2} and in probing both atomic and
spin correlations e.g. in spin-glasses \cite{murani} or magnetic 
defects in disordered alloys.  A full account of the applicability of the 
XYZ technique to this problem has been recently reviewed by 
Cywinski \textit{et al.}~\cite{cywinski}. Another
important application of the XYZ-method is that it allows to
separate incoherent and coherent atomic motions, as presented in
chapter~\ref{cohincoh}.

\begin{figure}[h]
\vspace{0 mm} \caption
   {Geometry of the XYZ polarization method. $\mathbf P_i$ and $\mathbf P_f$ are the polarizations
    of the incident and scattered beams, respectively. $\mathbf Q$ is the scattering vector.}
\label{xyz geometry}
\end{figure}

\subsection{Paramagnetic scattering}

According to the Rhodes and Wohlfarth \cite{rhodes} theory,
magnetic materials with $d$-electrons can be classified in
localized and itinerant systems. While for systems with localized
spin densities, the magnetic moment in the paramagnetic phase is
temperature independent, the ratio between paramagnetic and
ordered moments varies with temperature in the Stoner model. The
theory of spin fluctuations for localized and itinerant magnetic
systems is reasonably well developed in the paramagnetic phase
\cite{moriya}. In that respect, inelastic scattering of neutrons
provides direct experimental information on the spectrum of spin
fluctuations on an absolute scale as it gives access to the space-
and time-variation of the spin-spin correlation function
$S(\mathbf Q, \omega)$ that is related to the imaginary part of
the dynamical susceptibility $\chi(\mathbf Q,\omega)$
\cite{lovesey}. Paramagnetic scattering is usually very weak and
difficult to separate from coherent (phonons) and incoherent
nuclear scattering. However, the signal can be uniquely identified
in experiments by using the difference method, namely, the
difference between the spin-flip scattering as measured in a
(small) field parallel to $\mathbf Q$ and perpendicular to
$\mathbf Q$ contains only magnetic scattering \cite{squires,
ziebeck}. The reason being that, the
inelastic scattering from phonons is suppressed and nuclear
incoherent scattering as well as room background cancel. This
statement is generally valid as long as the nuclear magnetic
moments are disordered, i.e. $\langle I_x^2 \rangle=\langle I_y^2
\rangle =\langle I_z^2 \rangle = {1\over3}I(I+1)$, and $\langle
I_\alpha \rangle=0$. Nuclear ordering occurs only for extremely
low temperatures \cite{siemensmeyer}.

Therefore one obtains for example
\begin{equation}
 {1\over 2}{\biggl[{{d^2\sigma}\over{d\omega d\Omega}}\biggr]}_m =
 { \biggl[ {{d^2\sigma^{+,-}}\over{d\omega
 d\Omega}}\biggr]}_{\parallel}-
 {\biggl[{{d^2\sigma^{+,-}}\over{d\omega d\Omega}}\biggr]}_{\perp}
 ={ \biggl[{{d^2\sigma^{+,+}} \over {d\omega d\Omega}}\biggr]}_{\perp}
 -{\biggl[{{d^2\sigma^{+,+}}\over{d\omega d\Omega}}\biggr]}_{\parallel}. \label{paracs}
\end{equation}
Once the intensity of the paramagnetic fluctuations is measured,
the $E$-integrated intensity can be put on an absolute scale by
comparison with an acoustic phonon measured close to a Bragg peak
\cite{ishikawa} or by using an incoherent scatterer like vanadium.
Hence, an effective, paramagnetic moment can be found, defined by
\begin{equation}
M(q)={{1}\over{f(q)}}\biggl[\int_{0}^{\infty}{S(q,\omega)d\omega}\biggr]^{1/2},
\label{local moment}
\end{equation}
where $f(q)$ is the form factor \cite{ziebeck}. Ishikawa
\textit{et al.} \cite{ishikawa} pointed out that Eq.~\ref{local
moment} overestimates the amplitude of spin fluctuations at low
temperatures when the energy range of the spin fluctuations
extends beyond $\sim k_BT_c$. They propose instead to use the
Kramers-Kronig relation to obtain first the static susceptibility
that is linked via the fluctuation-dissipation theorem to the
amplitude of the spin fluctuations through $\langle
M^2(q)\rangle=3k_BT\chi(q)$ for $\hbar \omega \ll k_BT$ and
\begin{equation}
\chi(q)= g^2\mu_B^2\int_{-\infty}^{\infty}{{S(q,
\omega)[1-\exp(-\hbar \omega/k_BT)]}\over {\hbar \omega}}d\omega.
\label{kramers kronig}
\end{equation}
Measurements of the paramagnetic fluctuations in MnSi with a
coarse energy resolution (so that the $E$-integration is
automatically performed) with Eq. \ref{local moment}
\cite{ziebeck} and Eq. \ref{kramers kronig} \cite{ishikawa} show
that the amplitude of the local magnetic moment indeed increases
with increasing temperature, in agreement with self-consistent
renormalisation theory (Fig.~\ref{mnsi}) \cite{moriya}.

\begin{figure}[h]
\vspace{0 mm} \caption
   {Temperature dependence of $\rm 4\pi q^2 \langle M^2_q \rangle$ plotted against $q$ in MnSi
    as measured with polarized neutrons by Ishikawa \textit{et al.} \cite{ishikawa}. It is
    apparent from the figure that the mean-square amplitude of the spin fluctuations increases
    with increasing temperature in agreement with the calculations of the self-consistent
    renormalisation theory \cite{moriya}.}
\label{mnsi}
\end{figure}

Using the difference technique, the scaling behavior of many
different itinerant ferromagnets has been investigated in the
paramagnetic phase and it was shown that the scattering functions
of Fe and Ni can be modeled above $T_C$ by a simple Lorentzian
scattering function given by \cite{wicksted}
\begin{equation}
S(q, \omega)={{\omega}\over{1-\exp{(-\omega/T)}}}
\chi(q=0){\kappa^2\over\kappa^2+q^2}{{\Gamma_q}\over{\Gamma_q^2+\omega^2}},
\label{sqw}
\end{equation}
where the inverse correlation length
$\kappa=\kappa_0(T/T_c-1)^{0.7}$, the line-width $\Gamma_q = A
q^{2.5}f(\kappa/q)$, and $f(x)$ is approximately given by the
R\'esibois-Piette scaling function~\cite{resibois}.

To conclude, polarized neutron scattering is a powerful method to
measure paramagnetic fluctuations in particular when the signal is
weak.

\subsection{Transverse and longitudinal excitations in ferromagnets}

The magnetic properties of compounds with localized spin densities
are usually described by the Heisenberg Hamiltonian
\begin{equation}
H=-\sum_{i,j}{J_{ij}\mathbf S_i\cdot \mathbf S_j}
\label{heisenberg}
\end{equation}
where $J_{ij}$ is the exchange integral between the spins located
at the $i$- and $j$-position, respectively \cite{fazekas}.
Depending on the sign of the exchange integral, Eq.
\ref{heisenberg} favors either antiferromagnetic or ferromagnetic
ground-states. If exchange interactions extend beyond nearest
neighbors, competing effects can occur that may lead to
non-collinear or even incommensurate magnetic structures.

Because of its simplicity, the Heisenberg ferromagnet is often
taken as model system to study the properties of phase
transitions. Within the simple picture of localized spins, long
range order is lost due to the thermal excitation of spin waves
that evolve into the critical scattering at $T_C$. Using
unpolarized neutron scattering the spin dynamics close to $T_C$
has been investigated in detail \cite{passell}. It was shown that
the spin waves, i.e. the spin fluctuations transverse to the magnetisation
vector $\mathbf
M$, renormalise close to $T_C$ and that the susceptibility
$\chi(q)$ as measured at small angles diverges at small $\bf q$
for $T \rightarrow T_C$. Here, $\bf q$ is the reduced momentum
transfer with respect to the nearest Bragg peak. The divergence of
$\chi(q)$ is due to longitudinal fluctuations along $\mathbf M$ because the cross section for
spin waves does not contain a correlation length that diverges at
$T_C$. Unpolarized neutron scattering was not successful in
detecting the longitudinal fluctuations in ferromagnets in contrast
to the situation in antiferromagnets \cite{horn,coldea}, where
they can be easily observed.

The longitudinal fluctuations can be isolated by means of
inelastic neutron scattering with polarization analysis
\cite{boni91}. The experiment is performed by measuring the
differential spin-flip and non spin-flip cross sections from a
ferromagnetic sample, for example EuS, that is saturated in a
vertical magnetic field $\mathbf B_v$ that is perpendicular to the
scattering vector $\mathbf Q$. Fig.~\ref{EuS} shows three typical
measurements on EuS that have been performed longitudinal and
transverse to the reciprocal lattice point $(200)$ at $0.93 T_C$
\cite{boni95}.

\begin{figure}[h]
\vspace{0 mm} \caption
   {Constant-Q scans probing magnetic fluctuations in the ferromagnetic phase of EuS.
    The solid lines are fits to the data using Lorentzian spectral weight functions convoluted
    with the resolution function of the spectrometer IN14 at the ILL. The longitudinal spin
    waves are reduced in intensity due to the dipolar interactions. The parallel fluctuations
    are quasielastic.}
\label{EuS}
\end{figure}

The spin-flip data shows the spin waves with polarization vectors
$\delta \mathbf S$ transverse and parallel to the reduced momentum
transfer $\bf q$. The former are the Goldstone modes of the system
and diverge like $\chi_{sw}^T \propto 1/q^2$ (Table~\ref{table2}).
The longitudinal spin waves attain a mass \cite{fisher} due to the
dipolar interactions and do not diverge, $\chi_{sw}^L \propto
1/(q^2+q_D^2)$. Finally the non spin-flip data shows the
longitudinal fluctuations that are quasielastic and diverge like
$\chi_z \propto 1/(q^2+\kappa^2)$. Because the width $\Gamma_q$ of
$\chi_z(q,\omega)$ is comparable to the spin wave energy $E_q$ it
is clear why the longitudinal fluctuations escaped detection with
unpolarized neutron scattering. The results are in qualitative and
quantitative agreement with a coupled mode analysis
\cite{lovesey2} and mode-mode coupling theory \cite{schinz}.

\begin{table}[t] \caption{Transverse and longitudinal susceptibilities of
  a Heisenberg ferromagnet with dipolar interactions in the ordered
  and paramagnetic phases for different directions of
  the momentum-transfer $\mathbf q$ and accessible by
  one-dimensional polarization analysis
  (taken from Ref. \cite{boni95}). }
\label{table2}
\begin{center}
\begin{tabular}{l|cc}
\hline
    &$T<T_c$    &$T>T_c$\\
\hline \hline $\mathbf M \parallel \mathbf q$   &${2\over
q^2}+{1\over {q^2+\kappa^2_z+q^2_D}}$ &${2\over
{q^2+\kappa^2}}+{2\over q^2}+{1\over {q^2+\kappa^2_z+q^2_d}}$ \\
$\mathbf M \perp \mathbf q$   &${1\over q^2}+{1\over
{q^2+q^2_D}}+{1\over {q^2+\kappa^2_z}}$ &${2\over
{q^2+\kappa^2}}+{2\over q^2}+{1\over {q^2+\kappa^2_z+q^2_d}}$ \\
\hline
\end{tabular}
\end{center}
\end{table}

\subsection{Spin waves and phasons in incommensurate, antiferromagnetic Cr}

One of the outstanding features of antiferromagnetic Cr is the
occurrence of an incommensurate spin-density wave below $T_N =
311$ K that is transversely polarized ($\mathbf S$ perpendicular
$\bf Q^\pm$) with $\mathbf Q^\pm = (1 \pm \delta, 0, 0)$ being the
incommensurate wave-vector \cite{fawcett}. The magnetic
excitations exhibit many unusual features that are not well
understood. In particular, the magnetic modes that originate from
the magnetic satellite peaks at $\bf Q^\pm$ have such a steep
dispersion that the creation and annihilation peaks cannot be
resolved anymore. Using inelastic scattering of unpolarized
neutrons and analyzing the width of the peaks in constant energy
scans led to the conclusion that the velocity of the excitations
is $c_{sw} \simeq 1020$ meV\AA\ \cite{als-nielsen}. This value
deviates significantly from the theoretical value of a random
phase approximation (RPA) that is given by $c_{sw}^{th} =
\sqrt{1\over3} v_F \simeq 1500$ meV\AA, where $v_F$ is the Fermi
velocity \cite{fishman}. Using unpolarized neutron scattering it
has been shown that transverse (with respect to the staggered
magnetization) as well as longitudinal excitations contribute to
the inelastic scattering that emerges from the incommensurate $\bf
Q^\pm$ satellite peaks \cite{lorenzo}.

In the absence of sizable magnetic-nuclear interference contributions in
the cross-section, the most direct way to separate the transverse from the
longitudinal fluctuations is the use of longitudinal polarimetry. For
such an experiment it is necessary to use a Cr single-crystal
cooled through $T_N$ in a large magnetic field in order to induce
a single-$\mathbf Q$ state. During the experiment, a vertical
field ${\bf B} = 4$ T was applied along $[0 0 1]$ in order to
enforce a single-domain spin density wave with the magnetic
moments aligned along the $[0 1 0]$ direction. In this
configuration, the spin-flip scattering is due to the longitudinal
modes and the non spin-flip scattering due to the transverse
modes.

Fig.~\ref{Cr} shows constant energy scans for $E = 4.2$ meV
measured in the transverse spin-density-wave phase at $T = 230$ K
($0.74 T_N$) \cite{boni98}. It is clearly seen that the inelastic,
incommensurate peaks with transverse polarization are
significantly sharper than the corresponding longitudinal peaks.
Therefore, the mode velocity of the spin waves, $c_{sw}$, is
significantly larger than the mode velocity of the phason modes,
$c_{ph}$. These results are in qualitative agreement with results
of RPA theory \cite{fishman}. In addition, the data shows that the
enhanced magnetic scattering at $(1 0 0)$ and $E =4.2$ meV has a
longitudinal polarization. Without going into further details the
results indicate that a proper understanding of the magnetic
excitations in Cr can only be gained if polarization analysis is
used.

\begin{figure}[h]
\vspace{0 mm} \caption
   {Constant-$E$ scans at 4.2 meV, probing the longitudinal and transverse excitations
    along the $[100]$ direction in the transverse spin-density-wave phase of Cr at
    $T = 230$ K. The inset shows the intersection of the constant-$E$ scan with the dispersion
    of the the transverse (solid lines) and the longitudinal modes (broken lines) as well
    as the Fincher-Burke modes.}
\label{Cr}
\end{figure}

\subsection{Magnetic excitations in a heavy fermion superconductor}

The heavy fermion superconductor $\rm UPd_2Al_3$ exhibits the
unusual coexistence of antiferromagnetism and superconductivity
below $T_c=2$ K, i.e. the ordered magnetic moments of the
\textit{f}-electrons of $U$ persist in the superconducting phase
\cite{geibel}. This has been taken as a sign that the interplay of
magnetism and superconductivity could be studied in this material.
Neutron \cite{krimmel} and x-ray scattering \cite{gaulin}
experiments have shown that the magnetic structure of $\rm
UPd_2Al_3$ consists of ferromagnetic planes stacked along the
$c$-axis with a propagation vector $\rm \mathbf Q_0=(0,0,0.5)$.
The magnetic moments are confined within the hexagonal plane and
are found to have an unusually large value of $\rm \mu=0.85 \mu_B$
at saturation. First elastic \cite{krimmel2, kita} and inelastic
\cite{petersen} neutron scattering experiments could not
unambiguously reveal any change in the magnetic properties of $\rm
UPd_2Al_3$ upon cooling the sample below the superconducting
transition temperature.

Inelastic neutron scattering experiments \cite{sato, metoki}
performed with an improved energy-resolution as compared to the
work of Petersen et al. \cite{petersen} showed that there exist
two contributions to the spectrum of magnetic fluctuations in $\rm
UPd_2Al_3$. While the first one corresponds to the spin-wave
previously measured by Petersen et al. \cite{petersen}, a second
mode localized around the antiferromagnetic wave-vector $\mathbf
Q_0$ is observed in the energy range $0 < E < 0.5$ meV. This lower
energy mode is heavily damped for all temperatures in the
antiferromagnetically ordered phase and strongly sharpens upon
passing into the superconducting phase. At the lowest temperature
the low-energy mode develops an apparent energy gap with a value
comparable to $T_c$ \cite{metoki, bernhoeft}. Further evidence of
a strong interplay between magnetic fluctuations and
superconductivity in this compound originates from the use of
polarized neutrons as shown in Fig. \ref{UPdAl}. Using a polarized
beam,  it was possible to show that the two magnetic modes are
both polarized transverse to the magnetization vector and hence
are likely to interact with each other \cite{bernhoeft2}.

To perform this experiment the sample was field-cooled, so that
the magnetic domains could be aligned along the magnetization
vector $\mathbf M$. Using a neutron polarization perpendicular to
the scattering plane, it turns out
that magnetic fluctuations parallel to the magnetization are
non-spin flip, while those perpendicular to $\mathbf M$ appear in
the spin-flip channel. Analysis of the line-shape of the inelastic
neutron scattering data suggests that \textit{f}-electrons located
in a small energy range around the Fermi surface play a
significant role in forming the superconducting state in $\rm
UPd_2Al_3$ \cite{bernhoeft3}.

\begin{figure}[h]
\vspace{0 mm} \caption
   {Experimental data from $\rm UPd_2Al_3$ at the antiferromagnetic wave vector
    $\mathbf Q_0=(0, 0, 0.5)$ and $T=150$ mK. The data were taken with a fixed outgoing
    neutron wave vector of $k_f=1.15$ \AA$^{-1}$. Frame a) and b) are taken with polarized
    neutrons. For means of comparison a scan measured with  unpolarized neutrons
    is shown in frame c). In frame a) and b), the transverse response is shown
    as black symbols, whereas the longitudinal component is represented by open circles.
    See Ref.~\cite{bernhoeft2} for details.}
\label{UPdAl}
\end{figure}

\subsection{Magnons and Solitons in low-dimensional systems}

The magnetic properties of low-dimensional compounds have
attracted a lot of attention as new effects due to quantum
fluctuations are strong. For one-dimensional Heisenberg
antiferromagnets the ground states and energy excitations are
different for integer- and half-integer spins \cite{haldane}.
Antiferromagnetic chains with $S = 1/2$ spins have a disordered
ground-state. The low-lying excitations are characterized by a
continuum of excitations without energy gap at the zone center. On
the other hand, for integer-spins a finite energy gap was
predicted by Haldane and obtained by numerical calculations
\cite{haldane,affleck}. Examples of materials exhibiting a Haldane
gap are NENP \cite{regnault}, $\rm Y_2BaNiO_5$ \cite{darriet}, or
$\rm CsNiCl_3$ \cite{enderle}. The characteristics expected for a
Haldane system have been observed in these compounds by inelastic
neutron scattering, like i) a periodicity of $2\pi$ in the magnon
dispersion, ii) line-width broadening of the magnetic excitations
as a function of momentum transfer indicating the presence of a
two-magnon continuum, and iii) a large field dependence of the
magnetic excitations \cite{regnault2}.

Polarized neutron scattering experiments have shown that the
energy gap in the spectrum of magnetic excitations in $\rm
CsNiCl_3$ is a triplet \cite{steiner}. In such
quasi-one-dimensional antiferromagnets, weak inter-chain exchange
interactions $J'$ can lead to a N\'eel phase at low temperatures.
In fact the ordering temperature depends on the ratio of the
intra-chain interactions $J$ to $J'$. Interestingly, for such
systems, where antiferromagnetic ordering is close to disorder,
linear spin-wave theory does not account properly neither for the
energy-dependence of the magnetic excitations nor for the number
of magnetic modes. In particular, the existence of a
longitudinally polarized magnetic mode that cannot be predicted by
standard or modified spin-wave theory has been proven by means of
inelastic polarized neutron scattering in $\rm CsNiCl_3$
\cite{enderle} and $\rm Nd_2BaNiO_5$ (Fig.~\ref{Nd2BaNiO5})
\cite{raymond}. The instrumental set-up was chosen so that
magnetic excitations transverse to the magnetic moments could be
separated from the fluctuations along the spin direction. With
such a geometry, it could be shown that additional excitations
with longitudinal polarization are present in the spectrum of
$S(\mathbf Q, \omega)$, in agreement with calculations based on
renormalization-group theory \cite{affleck2}.

\begin{figure}[h]
\vspace{0 mm}
\caption
   {Temperature dependence of constant Q-scans measured in $\rm Nd_2BaNiO_5$
    with polarized neutrons. Open and black circles refer to spin-flip and non spin-flip
    scattering respectively (Taken from Ref. \cite{raymond}).  }
\label{Nd2BaNiO5}
\end{figure}

The quasi-one-dimensional $S = 1/2$ inorganic compound
$\rm CuGeO_3$ presents the particularity to undergo a chemical
phase transition below $T = 14$ K to a phase, called
\textit{Spin-Peierls} phase, where the copper chain is dimerised
\cite{pouget}. Consequently, the exchange interactions along the
chain direction are not uniform anymore but alternate with values
$J$ and $J'$, respectively. The magneto-elastic interaction is
presumably responsible for this phase transition characterized by
a non-magnetic ground state. For such a system, the spectrum of
magnetic excitations attains a gap at the zone center whereas the
first excited states are triplets. In $\rm CuGeO_3$ the gap has a
value of $\Delta \sim 2.5$ meV \cite{nishi} while away from the
zone center, the spectrum of magnetic fluctuations is strongly
dispersive along the copper chain direction. High-resolution
inelastic polarized-neutron experiments, however, revealed that
there is a second energy gap in this compound which separates the
low-energy magnon-like mode from a continuum of excitations
extending to higher energies \cite{lorenzo97}. The occurrence of
two energy gaps in the spectrum of magnetic excitations in $\rm
CuGeO_3$ is clearly a signature of strong quantum fluctuations in
$S = 1/2$ antiferromagnetic chains. This behavior differs
drastically from one-dimensional systems with large spin number
$S$ that in some cases can be described by the classical
sine-Gordon equation. The combined effects of non-linearity and
dispersion in these systems lead in addition to the linear
excitations to a special class of excitations called 'solitons'
\cite{steiner}.

In a ferromagnetic chain, an excitation of soliton-type can be viewed as
a $2\pi$ turn of the spins over a small distance, in a similar way to a domain
wall that would propagate through the crystal. These excitations
are accessible to inelastic neutron scattering and in particular
to polarized neutrons that allow to measure selectively the
different space and time correlation functions $S^{x,x}(\mathbf Q,
\omega)$, $S^{y,y}(\mathbf Q, \omega)$ and $S^{z,z}(\mathbf Q,
\omega)$ \cite{steiner}. Although experiments with unpolarized
neutrons have shown strong evidence of non-linear excitations in
the chain compounds $\rm CsNiF_3$ and TMMC, a crucial test for the
existence of solitonic excitations is the measurement of the
components fluctuating along and perpendicular to an applied
magnetic field \cite{steiner2}.

By separating the longitudinal part $S^{\parallel}(\mathbf Q,
\omega)$ of the dynamic structure factor from the transverse part
$S^{\perp}(\mathbf Q, \omega)$ with longitudinal polarimetry
analysis (see Fig. \ref{TMMC}), Boucher \textit{et al.}
\cite{boucher} were able to study the wave-vector and energy
dependence of solitonic fluctuations in the antiferromagnetic
chain compound $\rm (CD_3)_4NMnCl_3(TMMC)$. This study lead to the
result that amplitude and lifetime of the solitons are strongly
affected by collisions with magnons and by mutual interactions. As
a consequence, the line shape and the line width of the
experimental dynamical susceptibility differ from the actual
theoretical calculations based upon the low density
non-interacting soliton gas model \cite{steiner}.

\begin{figure}[h]
\vspace{0 mm} \caption
   {Spectra of magnetic excitations measured in TMMC with inelastic polarized neutron scattering
    showing that the transverse and longitudinal fluctuations are different. See text
    and Ref. \cite{boucher} for details.  }
\label{TMMC}
\end{figure}

\section{Self and collective diffusive atomic motions}
\label{cohincoh}

Collective motions of light atoms in metals consist of two
different processes \cite{sinha}. The first one can be viewed as
pure diffusion of ions through the lattice while the second
process involves cooperative hopping of mutually interacting
particles. Hence, the neutron scattering functions contain
incoherent and coherent scattering contributions that are given
within the random phase approximation by  Lorentzian functions centered
around zero energy transfer
\begin{eqnarray}
S_{inc}(Q, \omega)={1\over \pi}{{D_tQ^2}\over
{(D_tQ^2)^2+\omega^2}} \nonumber \\ S_{coh}(Q,
\omega)={{S(Q)}\over \pi}{{D_cQ^2}\over {(D_cQ^2)^2+\omega^2}}.
\label{diffusion}
\end{eqnarray}
$S(Q)$ is the static structure factor and $D_t$ and $D_c$
are the coefficients of incoherent and coherent diffusion,
respectively.

To separate the two quasi-elastic scattering processes which
appear simultaneously in the neutron spectrum, it is best to use
polarization analysis. In analogy to paramagnetic scattering,
coherent and incoherent processes can be isolated by calculating
the difference between non spin-flip and spin-flip scattering. In
the case of different isotopes and disordered nuclear spins, the
matrix elements for non spin-flip and spin-flip scattering are
given for the coherent cross-section
by
\begin{eqnarray}
 \sigma^{++}_{coherent}&=&\sigma^{--}=\langle N\rangle_{iso}^2 \\
 \sigma^{+-}_{coherent}&=&\sigma^{-+}=0
\end{eqnarray}
and for the incoherent scattering by \cite{squires}
\begin{eqnarray}
 \sigma^{++}_{incoherent}=\sigma^{--}_{incoherent}&=&\langle N^2\rangle_{iso}-\langle N\rangle^2_{iso}
 +{1\over 3} \langle B^2I(I+1) \rangle_{iso} \\
 \sigma^{+-}_{incoherent}=\sigma^{-+}_{incoherent}&=&{2\over 3} \langle B^2I(I+1) \rangle_{iso},
\end{eqnarray}
where $\langle \cdots\rangle_{iso}$ refers to isotopic averaging
and $I$ to the nuclear spin \cite{squires}. In compounds that
contain scatterers with one isotope only, the coherent
cross-section is obtained by dividing the spin-flip scattering by
$2$ and subtracting the result from the non-spin flip intensity.
In this case, all the incoherent scattering is spin-flip
scattering. Fig. \ref{alpha-NbD} shows the results of measurements
of the self and collective dynamics of deuterium in a single
crystal of Nb using the time-of-flight spectrometer D7 at ILL with
the method described in chapter~\ref{xyzxyz} \cite{cook}. It is
clear that polarization analysis gives an unambiguous separation
of the incoherent and coherent quasi-elastic signals over a large
range of momentum transfers $\bf Q$. Such a study allows to
determine distance and direction of jump processes, hopping and
residential times through the analysis of the $\bf Q$-dependence
and energy width of the Lorentzian functions in Eqs.
\ref{diffusion}.

\begin{figure}[h]
\vspace{0 mm} \caption
   {Coherent and incoherent scattering processes observed and calculated in
    $\alpha'$-NbD$_{0.7}$ by means of polarized
    neutron scattering on the multi-counter time-of-flight spectrometer D7 at the ILL
    (taken from Ref. \cite{cook}).}
\label{alpha-NbD}
\end{figure}

\section{Conclusions}

The examples given in the previous chapters have shown that
neutron scattering with polarized neutrons has become a very
important means to measure magnetic properties over a wide range
of $\bf Q$ and $\omega$ and to distinguish between coherent and
incoherent excitations in materials. Most experiments with
polarized neutrons are being performed up to now at sources
providing continuous neutron beams. The main reason being that
most neutron polarizers are more ideally suited for applications
with constant wavelength. The recent advances in the field of
supermirrors and $^3$He filters have improved the situation. The
new devices allow to extend polarization analysis to high neutron
energies and to use large area detectors. Therefore, we expect
that polarization analysis will also soon become a standard
technique at pulsed neutron sources.

Recently, new developments in the field of magnetism have emerged
that rely strongly on new developments in polarized neutron
scattering. As a first example, we mention systems that can be
characterized by sets of exponents, that differ according to the
space- and spin-dimensionality and hence can be grouped into
universality classes. In frustrated spin systems, the order
parameter includes a term describing the spin chirality $\mathbf
C=[\mathbf S_1 \times \mathbf S_2]$. A direct observation of the
fluctuations of the chiral variable is, however, impossible with
unpolarized neutron scattering as these are related to four-spin
correlation functions. Because the chiral part of the neutron
cross-section is polarization dependent \cite{maleyev} it can be
observed with polarization analysis.

As a second example, we mention that the Cryopad technique opens
the possibility to apply spherical neutron polarimetry in
inelastic neutron scattering as it allows to measure the
nuclear-magnetic interference term directly. Recently, Maleyev
\cite{maleyev99} has reconsidered the implication of the
nuclear-magnetic interference term (NMIT) for inelastic scattering
and has shown that in analogy with elastic scattering, it leads to
a dependence of the neutron cross-section upon $\mathbf P_i$,
namely to a finite polarization of the scattered neutrons and to a
rotation of the initial polarization. In particular, Maleyev has
shown that the part of the dynamical susceptibility due to the
NMIT is non-zero if there is a spin-lattice interaction
characterized by an axial vector, as it is the case e.g. for the
Dzialoshinskii-Moriya (DM) interaction $\mathbf D\cdot[ \mathbf
S_i \times \mathbf S_j]$. Indications of the importance of the
DM-interaction in the spin-lattice coupling and hence on its
accessibility through inelastic spherical neutron polarimetry
originates from recent inelastic experiments performed in the
spin-Peierls compound CuGeO$_3$ where a rotation of the final
polarization $\mathbf P_f$ has been detected when the incident
polarization is chosen parallel to the scattering vector
\cite{regnault3}. Such measurements have shown that spherical
neutron polarimetry can be applied in the field of inelastic
neutron scattering, although counting times are long due to poor
statistics. A further application may be the study of
magneto-elastic coupling that plays an important r\^ole in invar
alloys \cite{brown4}. Together with new theoretical interest on
such problems, it is probable that this method will contribute to
an improved understanding of phase transitions, where spin-lattice
interactions are important, as such effects cannot be studied by
standard polarization analysis.

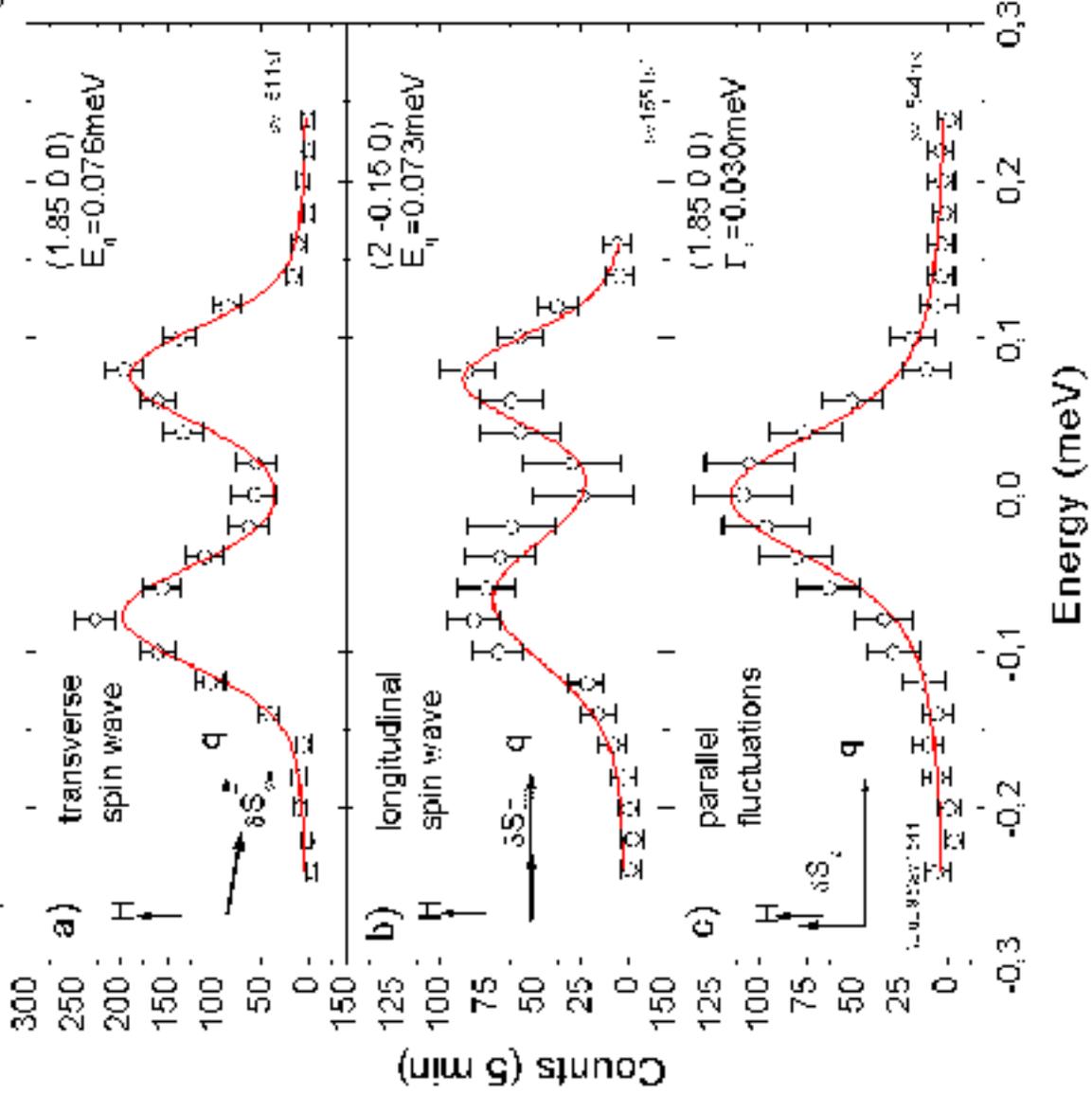
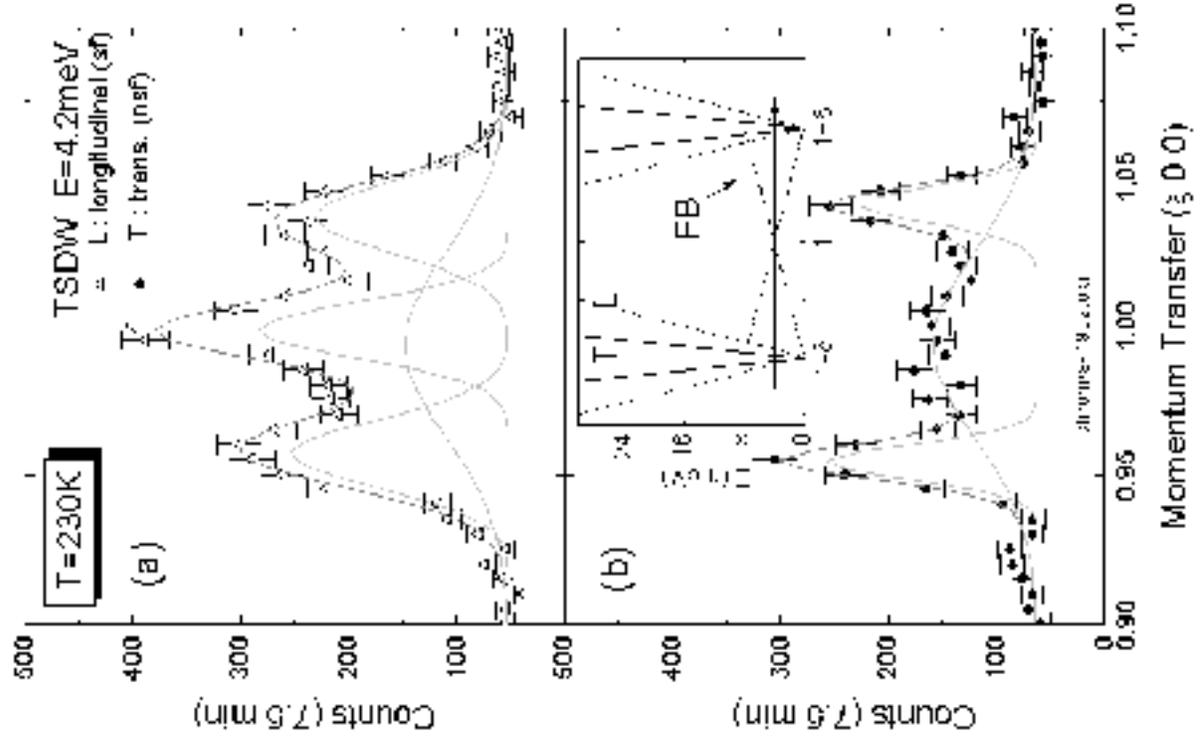
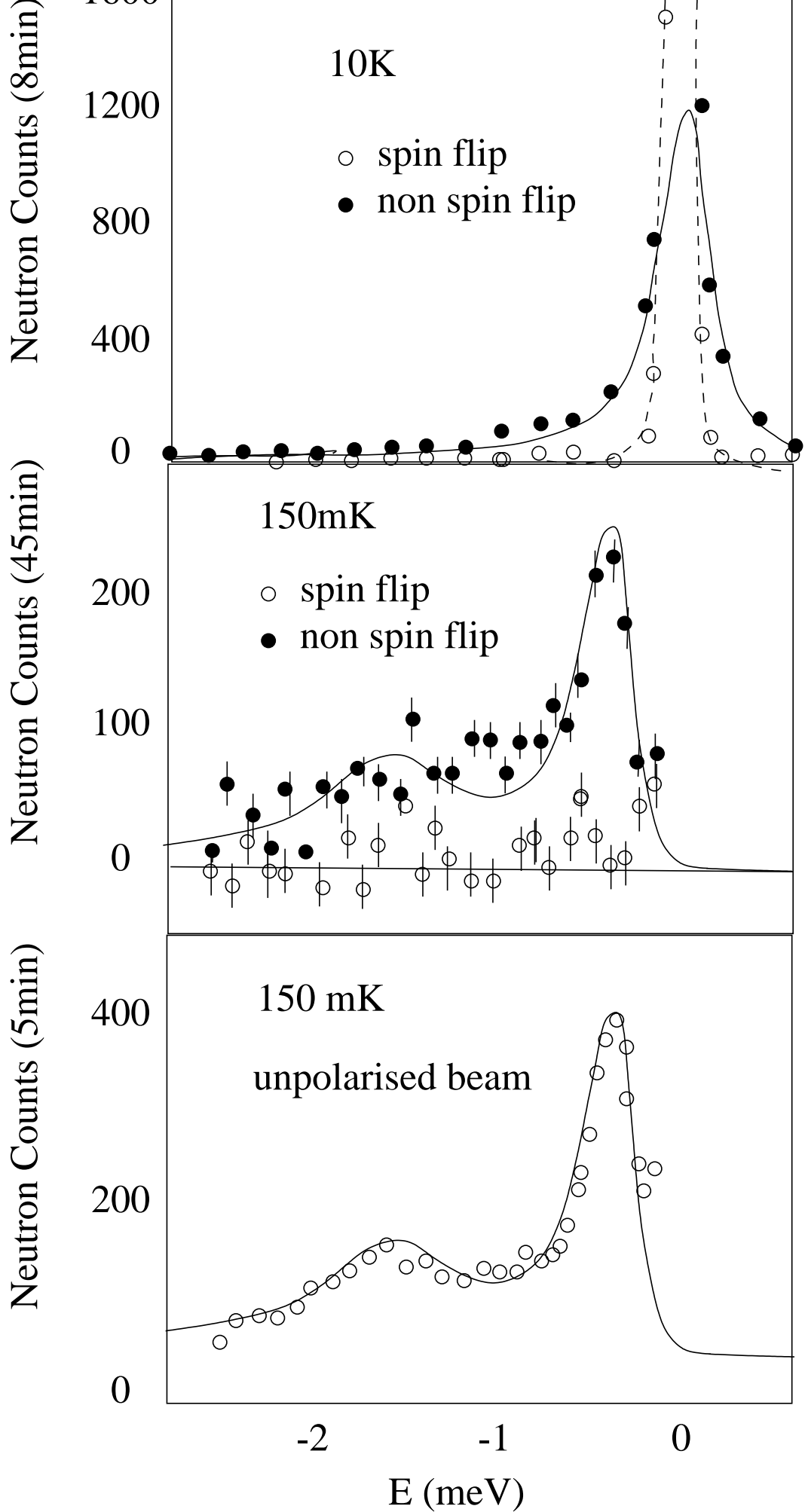
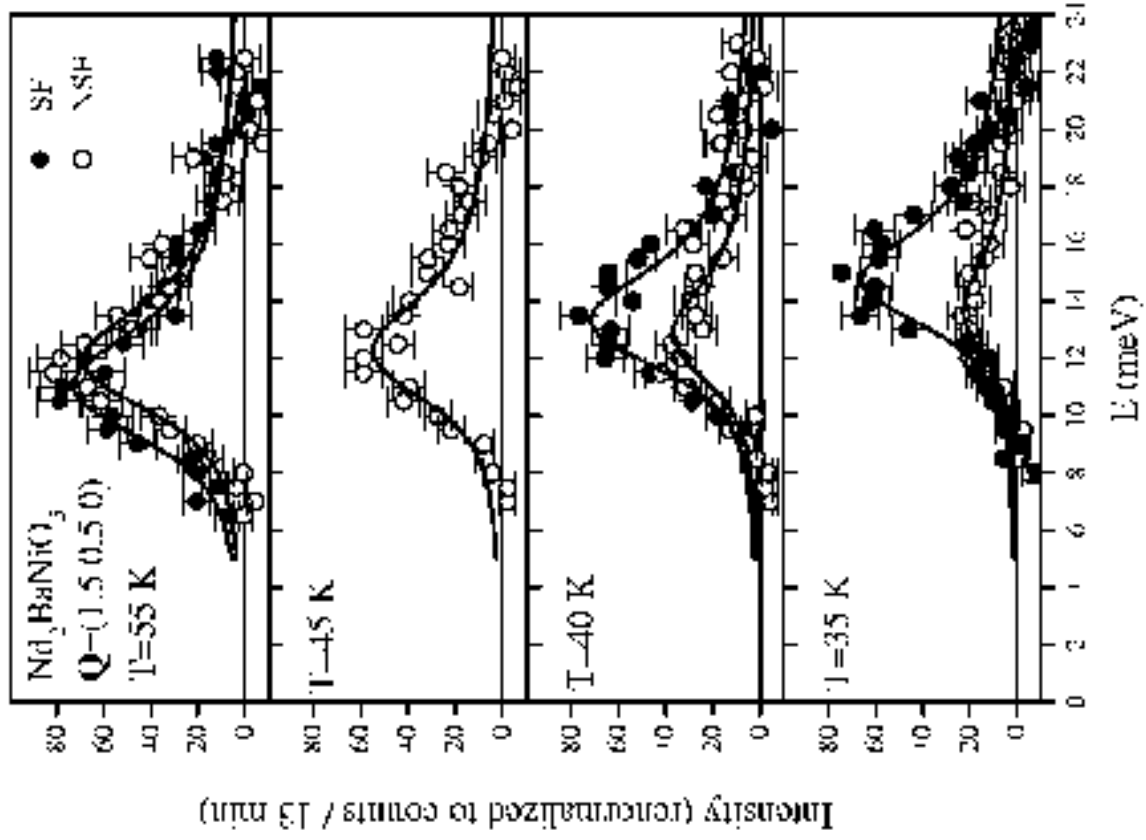
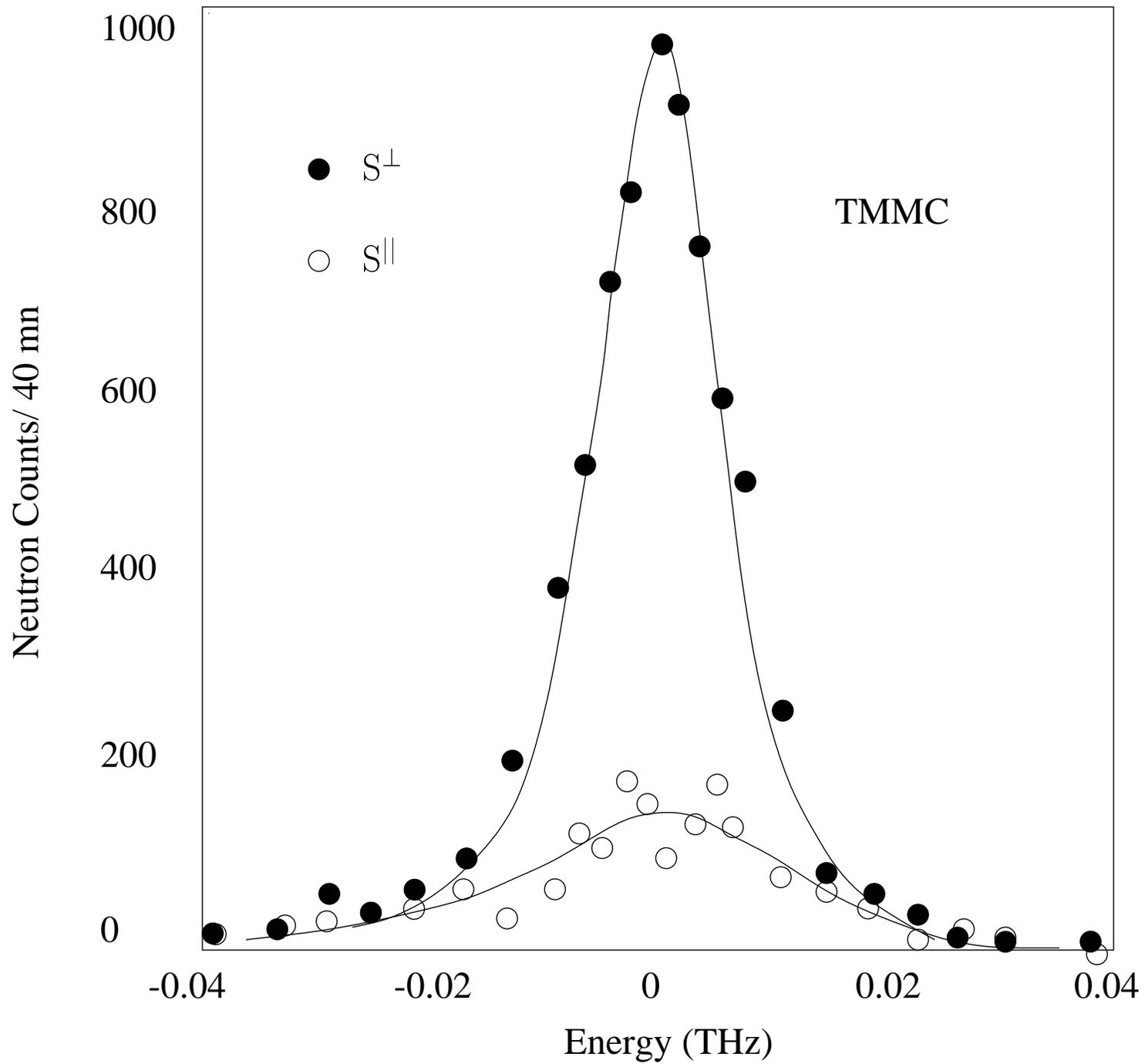
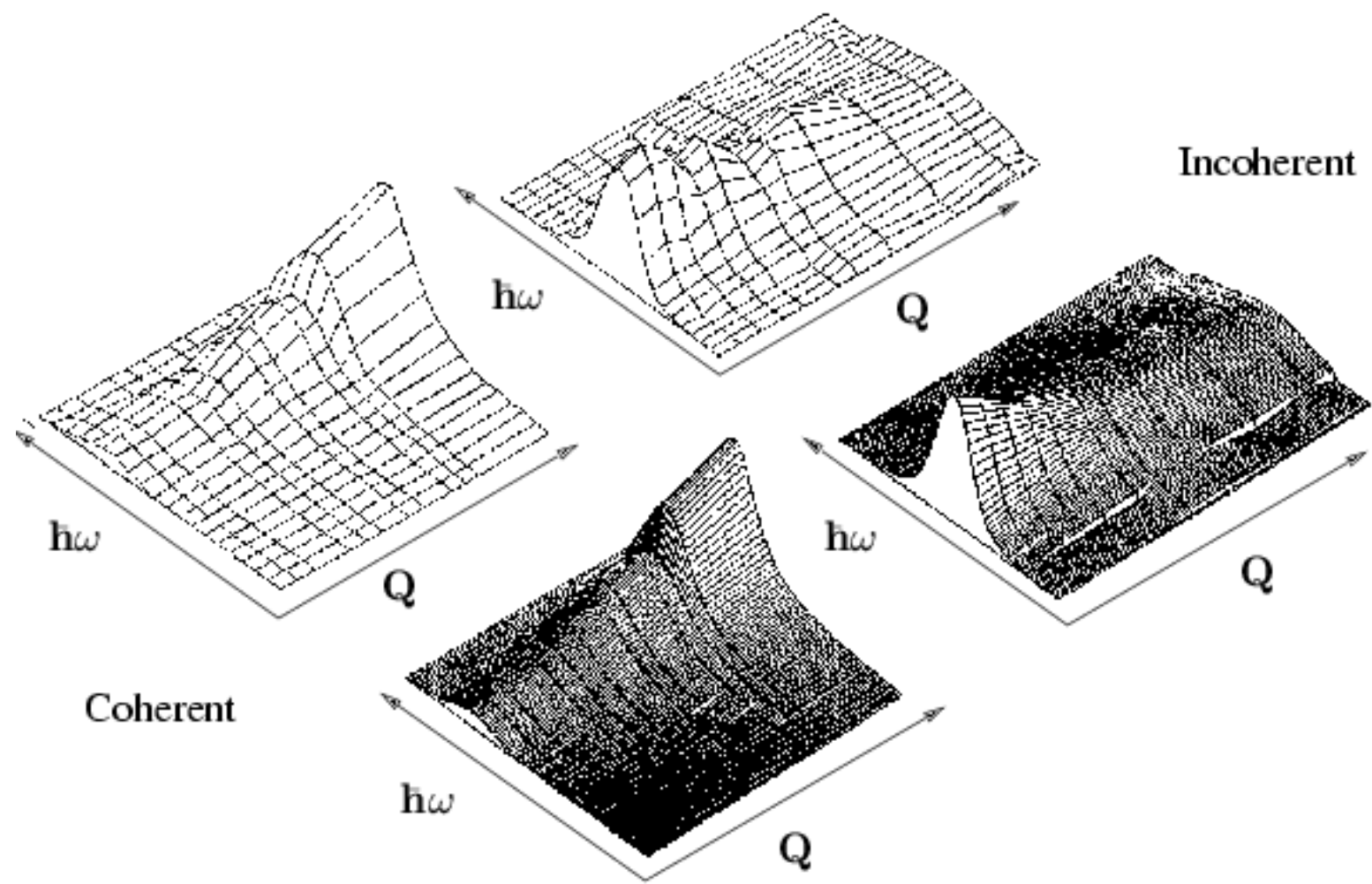
\end{document}